\documentclass[12pt]{article}

\usepackage[margin=1in]{geometry}

\usepackage{amsmath,amssymb,amsthm}
\usepackage{graphicx}
\usepackage[authoryear]{natbib}

\usepackage{threeparttable}
\usepackage{booktabs}
\usepackage{bm}
\usepackage{url}
\usepackage{hyperref}

\newcommand{\colrule}{\midrule}
\newcommand{\botrule}{\bottomrule}
\makeatletter
\long\def\tbl#1#2{\caption{#1}\centering #2}
\makeatother
\newenvironment{tabnote}{\par\vskip5pt\footnotesize}{\par}
\newenvironment{keywords}{\par\vskip11pt\noindent\textit{Keywords}: \ignorespaces}{\par}
\newenvironment{classcode}{\par\vskip6pt\noindent\textit{JEL Classification}: \ignorespaces}{\par\vskip12pt}

\theoremstyle{plain}
\newtheorem{theorem}{Theorem}[section]

\newtheorem{proposition}[theorem]{Proposition}

\theoremstyle{definition}

\theoremstyle{remark}

\begin{document}

\title{Optimal Signal Extraction from Order Flow: A Matched Filter Perspective on Normalization and Market Microstructure}
\author{Sungwoo Kang\thanks{Corresponding author. Email: krml919@korea.ac.kr}\\
\textit{Department of Electrical and Computer Engineering}\\
\textit{Korea University, Republic of Korea}}
\date{}

\maketitle

\begin{abstract}
We establish a general matched filter principle for order flow normalization: optimal normalization must match the scaling behaviour of the signal-generating process. For capacity-constrained institutional investors, market capitalization normalization ($S^{MC}$) is the matched filter; for volume-targeting traders (e.g., VWAP/TWAP algorithms), trading value normalization ($S^{TV}$) is optimal. Monte Carlo simulations confirm this principle works bidirectionally, with matched filters achieving up to $1.99\times$ higher signal correlation. Empirical validation using 2.7 million stock-day observations from the Korean market (2020--2024) reveals symmetric normalization dominance across investor types: domestic institutional flows predict next-day returns significantly under $S^{MC}$ ($t = 9.65$), while foreign flows exhibit stronger predictability under $S^{TV}$ ($t = 16.35$)---with no sign reversal at longer horizons, indicating durable private information rather than temporary price impact. These findings motivate the ``Informed Executor'' hypothesis: sophisticated foreign investors possess genuine private information but employ volume-targeting algorithms for stealth execution---volume-scaling reflects execution methodology, not absence of information. Information-theoretic validation using KL divergence independently corroborates these results. The matched filter principle generalises to any market where signal scaling varies across trader types, with implications for trading algorithms, factor construction, and market microstructure methodology.
\end{abstract}


\begin{classcode}G12, G14, C58\end{classcode}

\section{Introduction}

The measurement of trading intensity is central to modern finance. Whether estimating the probability of informed trading \citep{Easley1996}, constructing liquidity factors \citep{PastorStambaugh2003, AMIHUD200231}, or designing execution algorithms \citep{Almgren2003}, researchers and practitioners must normalize raw order flow---divide net buying (in dollars) by some measure of firm size or activity. This step is treated as routine, yet a careful survey of the literature reveals a striking and unexplained pattern: studies analysing informed institutional trading tend to normalize by market capitalization or shares outstanding \citep{CAMPBELL200966, Beber2011, LEWELLEN201162}, while studies of liquidity, market impact, and flow toxicity normalize by trading volume \citep{AMIHUD200231, Easley2012}. No systematic framework exists for determining when each normalization is appropriate, or for quantifying the consequences of choosing incorrectly.

This paper fills that gap. We demonstrate that normalization choice has first-order consequences for signal quality, and that the direction of the effect depends on the scaling behaviour of the trader class under study. The core problem is one of \textit{signal--noise separation}. When capacity-constrained institutional investors---pension funds, mutual funds, insurance companies---trade on private information, they scale their positions relative to firm market capitalization, reflecting capital allocation rules and capacity constraints. Noise traders, by contrast, respond to daily liquidity conditions, generating order flow that scales with trading volume. Normalizing institutional order flow by volume inadvertently scales the true information signal by the inverse of turnover---a highly variable, firm-specific quantity---creating systematic heteroskedasticity that obscures the relationship between flow and future returns.

This produces what we term the \textit{Participation Rate Fallacy}\label{subsec:prf}: a systematic distortion in which changes in trading volume are misattributed as changes in trading conviction. When volume drops, the denominator shrinks and the normalized signal inflates, mimicking a high-conviction trade even when the underlying position is unchanged; conversely, a volume surge dilutes a genuinely informed position into apparent inactivity. The distortion is especially pernicious because turnover itself proxies for noise rather than information. \citet{Banerjee2010} show theoretically that high trading volume often reflects differences of opinion rather than information arrival, and \citet{Barinov2014} demonstrates empirically that turnover captures firm-specific uncertainty and investor disagreement. Normalizing by a noise-contaminated denominator mechanically down-weights informed signals during periods of high disagreement---precisely when they may be most valuable---and up-weights them during low-activity episodes that may contain no signal at all.

We address this problem by reframing order flow normalization as a \textit{signal extraction} problem, drawing on matched filter theory from communications engineering \citep{Turin1960}. A matched filter maximises the signal-to-noise ratio by weighting the received signal according to the known structure of the transmitted signal. We establish a \textbf{general matched filter principle}: optimal normalization must \textit{match} the scaling behaviour of the signal-generating process. For capacity-constrained informed traders whose positions scale with market capitalization ($Q \propto M$), the matched filter is market capitalization normalization ($S^{MC} = D/M$), which recovers the pure information signal. For volume-targeting traders whose execution scales with trading volume ($Q \propto V$)---such as VWAP/TWAP algorithms \citep{Almgren2003}---the matched filter is trading value normalization ($S^{TV} = D/V$), which recovers the target participation rate. In each case, the mismatched filter introduces heteroskedastic distortion that corrupts the signal. This principle is \textit{bidirectional}: it predicts not only that $S^{MC}$ dominates for capacity-scaled flows, but symmetrically that $S^{TV}$ dominates for volume-scaled flows.

This paper makes four contributions. First, we develop a theoretical framework formalising the normalization problem as signal extraction, deriving the general matched filter principle from Jensen's inequality and SNR maximization (Section~\ref{sec:theory}). The framework generates sharp, testable predictions: the direction of normalization dominance should \textit{reverse} across investor types whose scaling behaviours differ. Second, we validate these predictions through 1,000 Monte Carlo simulations with 500 stocks each (Section~\ref{sec:montecarlo}). Market capitalization normalization achieves $1.32\times$ higher correlation when signals are capacity-scaled, while trading value normalization achieves $1.13\times$ higher correlation when signals are volume-scaled---with both advantages increasing monotonically with turnover heterogeneity (up to $1.99\times$). Third, we provide comprehensive empirical validation using up to 2.7 million stock-day observations from the Korean equity market, 2020--2024 (Section~\ref{sec:empirical}). The Korean market offers a unique advantage: the Korea Exchange mandates daily disclosure of trading activity disaggregated by investor type---domestic institutional, foreign, and individual---enabling direct observation of order flow heterogeneity without reliance on classification algorithms. For domestic institutional flows, $S^{MC}$ achieves highly significant next-day return predictability ($t = 9.65$), while $S^{TV}$ is only marginally significant ($t = 2.10$); in the horse race, $S^{TV}$ reverses sign, confirming the Participation Rate Fallacy empirically. For foreign institutional flows, the pattern reverses as the theory predicts: $S^{TV}$ achieves $t = 16.35$, confirming the bidirectional nature of the matched filter principle. Horizon analysis reveals that domestic flows show partial sign reversal at 20 days (consistent with transient price impact), while foreign flows remain positive without reversal (indicating durable private information). An information-theoretic analysis using KL divergence independently corroborates these findings from a distributional perspective (Section~\ref{sec:information_theory}).

Fourth, and most unexpectedly, the empirical results motivate what we term the \textbf{Informed Executor Hypothesis}. Foreign institutional investors exhibit volume-scaled execution---consistent with algorithmic VWAP/TWAP strategies---yet their flows carry the strongest return predictability of any investor-signal combination ($t = 16.35$ vs.\ $9.78$ for domestic institutions). That their matched filter is $S^{TV}$ confirms the theory; what is surprising is the magnitude and persistence of the signal. We argue that sophisticated foreign investors possess genuine private information---from global research networks, cross-market signals, and early access to international capital flows---but deliberately employ volume-targeting algorithms to minimise their market footprint. Volume-scaling reflects their \textit{execution methodology}, not their \textit{information content}. This finding challenges the common assumption that volume-scaled trading indicates uninformed execution, and has implications for how researchers interpret algorithmic order flow in market microstructure studies.

The remainder of this paper is organised as follows. Section~\ref{sec:literature} reviews related literature. Section~\ref{sec:theory} develops our theoretical model and the general matched filter principle, covering both capacity-scaled and volume-scaled signals. Section~\ref{sec:montecarlo} presents Monte Carlo validation of the symmetric predictions. Section~\ref{sec:empirical} provides empirical evidence from the Korean stock market, where domestic institutions (capacity-scaling) and foreign investors (volume-scaling) exhibit the predicted divergent patterns. Section~\ref{sec:information_theory} provides complementary evidence from an information-theoretic perspective using Kullback-Leibler divergence. Section~\ref{sec:discussion} discusses implications, and Section~\ref{sec:conclusion} concludes.

\section{Literature Review}
\label{sec:literature}

Our work connects to several strands of research. While explicit application of matched filter theory to order flow normalization is novel, the existing literature provides the building blocks---models of informed trading and price discovery, empirical studies of institutional order flow, and the growing literature on algorithmic execution---that together motivate and contextualise our framework.

\subsection{Market Microstructure and Price Discovery}

The foundational models of \citet{Kyle1985} and \citet{GLOSTEN198571} establish that order flow conveys private information to market makers, with the price impact parameter reflecting the signal-to-noise ratio in aggregate flow. \citet{Hasbrouck1991} develops methods for decomposing price changes into permanent (information) and transitory (noise) components. These models distinguish informed from uninformed trading but do not address how researchers should \textit{normalize} observed flow to extract the embedded signal---the question at the center of our analysis. The ``Square Root Law'' of market impact, empirically documented by \citet{Torre1997} and related to the theory of power-law fluctuations from large institutional trades \citep{Gabaix2003}, states that price impact scales as $I \propto \sqrt{Q/V}$, which is often cited to justify volume normalization. However, this law describes the \textit{cost} of trading---the friction the market imposes---rather than the \textit{signal} embedded in informed positions. Our framework makes this distinction explicit: volume normalization is appropriate for estimating execution cost, but market capitalization normalization better extracts the informational signal from capacity-constrained traders.

\subsection{Institutional Order Flow and Normalization Practices}

Two influential studies illustrate the implicit normalization choices in the literature. \citet{CAMPBELL200966} infer institutional order flow from TAQ data and validate their measure against quarterly 13F filings, which report holdings as a percentage of shares outstanding---a market-capitalization-based metric. Their methodology naturally aligns with our $S^{MC}$ measure. \citet{Beber2011} define ``active sector order flow'' as the deviation from a cap-weighted baseline ($w_i = M_i / \sum M_j$), explicitly treating market capitalization as the neutral benchmark for institutional positioning. \citet{LEWELLEN201162} finds that institutional capital allocation scales with firm size rather than daily trading activity. These studies implicitly adopt the normalization our theory prescribes for capacity-constrained traders, though without formalising the underlying principle.

In contrast, the liquidity and flow toxicity literature normalizes by volume. \citet{AMIHUD200231} measures illiquidity as the ratio of absolute return to dollar volume, while \citet{PastorStambaugh2003} construct a liquidity factor from the return-volume relationship. The PIN framework \citep{Easley1996} and its volume-synchronised extension VPIN \citep{Easley2012} sample order flow in volume-time, measuring imbalance per unit of volume. While effective for their intended purposes---liquidity estimation and toxicity detection---these volume-normalized measures may not be optimal for alpha prediction when the underlying information signal scales with capacity. Our matched filter framework provides the theoretical basis for this distinction and formalises when each normalization is appropriate.

The turnover literature further supports our framework. \citet{DATAR1998203} document a negative relationship between turnover and expected returns, initially attributed to liquidity effects. Subsequent research reinterprets this finding: \citet{Banerjee2010} shows that high volume often reflects differences of opinion rather than information, and \citet{Barinov2014} argues that turnover proxies for firm-specific uncertainty and disagreement. These findings support our model specification in which noise trader flow scales with volume ($Q_{noise} \propto V$), providing the economic foundation for why volume normalization introduces distortion when applied to capacity-constrained informed flows.

\subsection{Algorithmic Execution and Volume-Scaled Trading}

The rise of algorithmic trading creates a distinct trader class whose order flow scales with volume rather than market capitalization, providing the empirical motivation for Case~2 of our framework. \citet{Almgren2003} develop optimal execution models where large institutional orders are spread across time to achieve target participation rates: $Q_{algo,i,t} = \eta \cdot V_{i,t}$. VWAP and TWAP algorithms implement this principle by splitting orders into smaller pieces proportional to daily volume. \citet{Hendershott2011} document the prevalence of algorithmic trading across modern markets. Research on foreign institutional trading in emerging markets \citep{Choe1999, Choe2005} documents that foreign institutions execute differently from domestic participants, consistent with algorithmic participation-rate targeting. For these volume-targeting traders, our theory predicts that trading value normalization ($S^{TV}$) is the appropriate matched filter---a prediction we validate empirically in Section~\ref{sec:empirical}.

\subsection{Signal Processing in Finance}

Our theoretical approach draws on matched filter theory from communications engineering \citep{Turin1960}, which establishes that detecting a known signal waveform in noise requires correlating the received data with a template matching the expected signal structure. While signal processing techniques have been applied to financial time series---\citet{Campbell1997} use Kalman filters and state-space models for temporal noise reduction---applications to \textit{cross-sectional} normalization problems are, to our knowledge, absent from the literature. Our contribution extends matched filter logic from the temporal to the cross-sectional domain, establishing that the optimal normalizer is the ``replica'' of the signal-generating process: $1/M_i$ for capacity-constrained traders (recovering the latent signal $\alpha_i$) and $1/V_i$ for volume-targeting traders (recovering the participation rate $\eta_i$).

\subsection{Summary of Literature Position}

Table~\ref{tab:literature_comparison} summarises how normalization approaches in the literature align with our matched filter perspective. The key observation is that papers studying informed institutional trading tend to normalize by market capitalization (or shares outstanding), while papers focused on liquidity and market impact normalize by volume. Our contribution provides a theoretical foundation for this implicit pattern via the general matched filter principle: optimal normalization matches the scaling behaviour of the signal source---market capitalization for capacity-constrained traders, trading volume for participation-rate traders.

\begin{table}
\begin{center}
\begin{minipage}{155mm}
\tbl{Literature Comparison: Normalization Approaches in Prior Research}
{\begin{tabular}{@{}p{2.5cm}p{3.5cm}p{2.5cm}p{3.5cm}p{3cm}@{}}\toprule
Research Area & Representative Papers & Normalization & Key Insight & Our Perspective \\\colrule
Market Microstructure & Kyle (1985); Glosten \& Milgrom (1985) & Order Flow / $\lambda$ & Price impact $\lambda$ scales with signal-to-noise ratio & Conflates execution cost with signal intent \\
Institutional Trading & Campbell et al.\ (2009); Beber et al.\ (2011) & Flow / Market Cap & Validates against 13F (cap-weighted) & Directly supports $S^{MC}$ \\
Liquidity Research & Amihud (2002); Pastor \& Stambaugh (2003) & Return / Volume & Captures illiquidity premium & Appropriate for cost, not alpha \\
Turnover Studies & Barinov (2014); Datar et al.\ (1998) & Volume / Shares Out & Turnover proxies disagreement/noise & Supports $Q_{noise} \propto V$ \\
Flow Toxicity & Easley et al.\ (2012) & Imbalance / Volume & VPIN detects adverse selection & Risk metric, not alpha signal \\
Our Approach & This paper & Flow / Matched Filter & $S^{MC}$ for capacity-scaled; $S^{TV}$ for volume-scaled & General matched filter principle \\\botrule
\end{tabular}}
\label{tab:literature_comparison}
\end{minipage}
\end{center}
\end{table}

\section{Theoretical Model}
\label{sec:theory}

\subsection{Setup and Primitives}

Building on the insight that different trader classes exhibit distinct scaling behaviours, we now formalise the normalization problem as a signal extraction problem.

Consider a cross-section of $N$ stocks indexed by $i \in \{1,\ldots,N\}$. Each stock has a market capitalization $M_i$ (known, time-invariant for simplicity) and daily traded value $V_i = \tau_i M_i$ (in currency units), where $\tau_i$ is the turnover rate. The true information content $\alpha_i$ is latent with mean zero and variance $\sigma_\alpha^2$. Future returns are given by $R_i = \gamma \alpha_i + \epsilon_i$, where $\epsilon_i \sim N(0,\sigma_\epsilon^2)$.

\subsection{Order Flow Generation}

Two types of traders generate order flow:

\textbf{Informed Traders} observe $\alpha_i$ and trade to exploit it. From mean-variance optimisation, optimal position size is proportional to expected return and inversely proportional to risk. Crucially, informed traders scale their absolute dollar positions by market cap, reflecting capacity constraints and capital allocation rules:
\begin{equation}
Q_{inf,i} = k \cdot \alpha_i \cdot M_i
\label{eq:informed}
\end{equation}
where $k > 0$ captures risk aversion and institutional constraints. We note that this model isolates the permanent information component for analytical clarity. In practice, real-world ``informed'' order flow is a mixture of permanent fundamental information and transient demand impact generated as markets absorb institutional positions. The matched filter extracts the total signal (both components); our horizon analysis (Section~\ref{subsec:investor_heterogeneity}) then decomposes it by examining whether predictability persists or reverses at longer horizons, separating durable information from temporary price pressure.

\textbf{Noise Traders} trade for non-informational reasons (liquidity needs, attention, behavioural biases). Their order flow scales with daily trading activity:
\begin{equation}
Q_{noise,i} = \zeta_i \cdot V_i
\label{eq:noise}
\end{equation}
where $\zeta_i \sim N(0,\sigma_\zeta^2)$ is independent of $\alpha_i$.

\textbf{Execution Traders} (e.g., algorithmic VWAP/TWAP strategies, index rebalancers) target participation rates rather than absolute positions. Their order flow scales with volume:
\begin{equation}
Q_{exec,i} = k \cdot \eta_i \cdot V_i
\label{eq:execution}
\end{equation}
where $\eta_i \sim N(0,\sigma_\eta^2)$ represents the signed deviation in participation rate from the market-neutral baseline.

\textbf{Observed Order Flow} depends on which trader type generates the signal of interest. For informed traders (Case 1):
\begin{equation}
D_i^{\text{Case1}} = k\alpha_i M_i + \zeta_i V_i
\label{eq:total_flow}
\end{equation}
For execution traders (Case 2):
\begin{equation}
D_i^{\text{Case2}} = k\eta_i V_i + \xi_i M_i
\label{eq:total_flow_exec}
\end{equation}
where $\xi_i \sim N(0,\sigma_\xi^2)$ represents capacity-based noise from fundamental traders (i.e., informed trading flow from Case~1, which is ``noise'' from the perspective of extracting execution signals).

\subsection{Signal Extraction Problem}

Our goal: extract the latent signal from observed $D_i$ to predict $R_i$. The optimal normalization depends on \textit{which trader type} generates the signal of interest.

\subsubsection{Case 1: Informed Traders (Capacity-Scaled Signals)}

For informed trader flow $D_i = k\alpha_i M_i + \zeta_i V_i$, where returns $R_i = \gamma\alpha_i + \epsilon_i$:

\textbf{Trading Value Normalization}:
\begin{align}
S_i^{TV} &= \frac{D_i}{V_i} = k\alpha_i \tau_i^{-1} + \zeta_i
\label{eq:tv_norm}
\end{align}
\textit{Problem}: The signal $\alpha_i$ is multiplied by $1/\tau_i$, creating heteroskedastic distortion.

\textbf{Market Capitalization Normalization}:
\begin{align}
S_i^{MC} &= \frac{D_i}{M_i} = k\alpha_i + \zeta_i \tau_i
\label{eq:mc_norm}
\end{align}
\textit{Advantage}: The signal $k\alpha_i$ appears unscaled. \textbf{$S^{MC}$ is the matched filter for Case 1.}

\subsubsection{Case 2: Execution Traders (Volume-Scaled Signals)}

For execution trader flow $D_i = k\eta_i V_i + \xi_i M_i$, where returns $R_i = \gamma\eta_i + \epsilon_i$:

\textbf{Trading Value Normalization}:
\begin{align}
S_i^{TV} &= \frac{D_i}{V_i} = k\eta_i + \xi_i \tau_i^{-1}
\label{eq:tv_norm_exec}
\end{align}
\textit{Advantage}: The signal $k\eta_i$ appears unscaled. \textbf{$S^{TV}$ is the matched filter for Case 2.}

\textbf{Market Capitalization Normalization}:
\begin{align}
S_i^{MC} &= \frac{D_i}{M_i} = k\eta_i \tau_i + \xi_i
\label{eq:mc_norm_exec}
\end{align}
\textit{Problem}: The signal $\eta_i$ is multiplied by $\tau_i$, creating heteroskedastic distortion.

\subsection{General Matched Filter Principle}

\begin{proposition}[General Matched Filter Optimality]
Let $\rho(S,R)$ denote the correlation between a normalized signal $S$ and future returns. The optimal normalization \textbf{matches} the scaling behaviour of the signal-generating process:
\begin{enumerate}
\item[(i)] \textbf{Capacity-scaled signals} ($Q \propto M$): $E[\rho(S^{MC},R)] > E[\rho(S^{TV},R)]$
\item[(ii)] \textbf{Volume-scaled signals} ($Q \propto V$): $E[\rho(S^{TV},R)] > E[\rho(S^{MC},R)]$
\end{enumerate}
whenever turnover $\tau_i$ exhibits cross-sectional dispersion and the signal-to-noise ratio is non-negligible (i.e., $\sigma_\alpha^2 / \sigma_\zeta^2$ is not vanishingly small). When the signal is negligible relative to noise, both normalizations are dominated by noise-floor effects, and the matched filter advantage may not hold (see table~\ref{tab:robustness}, Panel~A, $\sigma_\alpha = 0.01$).
\end{proposition}

\begin{proof}
\textbf{Case 1 (Informed Traders):} From equations \eqref{eq:tv_norm} and \eqref{eq:mc_norm}, when signal $\alpha$ scales with $M$:
\begin{align*}
S^{MC} &= k\alpha_i + \zeta_i \tau_i \quad \text{(signal unscaled)}\\
S^{TV} &= k\alpha_i \tau_i^{-1} + \zeta_i \quad \text{(signal distorted by $\tau^{-1}$)}
\end{align*}
In $S^{TV}$, the signal $k\alpha_i \tau_i^{-1}$ is also inflated, but this inflation increases signal \textit{variance} faster than it increases the signal's \textit{covariance} with returns: $\text{Cov}(k\alpha_i \tau_i^{-1}, \gamma\alpha_i) \propto E[\tau_i^{-1}]$ while $\text{Var}(k\alpha_i \tau_i^{-1}) \propto E[\tau_i^{-2}]$, and Jensen's inequality on the convex function $f(x)=1/x^2$ ensures $E[\tau^{-2}] > (E[\tau^{-1}])^2$. The net effect reduces correlation, yielding $\text{SNR}_{MC} > \text{SNR}_{TV}$.

\textbf{Case 2 (Execution Traders):} From equations \eqref{eq:tv_norm_exec} and \eqref{eq:mc_norm_exec}, when signal $\eta$ scales with $V$:
\begin{align*}
S^{TV} &= k\eta_i + \xi_i \tau_i^{-1} \quad \text{(signal unscaled)}\\
S^{MC} &= k\eta_i \tau_i + \xi_i \quad \text{(signal distorted by $\tau$)}
\end{align*}
By Jensen's inequality for convex $g(x)=x^2$ (reflecting that signal power scales with the square of the linear distortion), the variance inflation in $S^{MC}$ exceeds the noise scaling in $S^{TV}$. Specifically, in $S^{TV}$ the signal $k\eta_i$ passes through undistorted while noise is attenuated by $E[\tau^{-1}] < 1$; in $S^{MC}$ the signal covariance scales with $E[\tau_i]$ but signal variance scales with $E[\tau_i^2] > (E[\tau_i])^2$, inflating dispersion faster than predictive power. Therefore $\text{SNR}_{TV} > \text{SNR}_{MC}$.

The complete mathematical derivations are provided in Appendix~\ref{app:snr_derivation}.
\end{proof}

The SNR advantage of the matched filter has a direct information-theoretic analog: the Kullback-Leibler divergence between return distributions conditional on buying versus selling quantifies distributional separation---a measure of ``information distance'' that parallels our correlation-based SNR. Just as the matched filter maximises SNR by preserving signal structure, it also maximises distributional separation by avoiding the heteroskedastic distortion that contaminates the mismatched filter. Section~\ref{sec:information_theory} validates our matched filter results from this complementary information geometry perspective.

\subsection{Economic Interpretation}

The matched filter principle reflects fundamental differences in trader behaviour:

\textbf{Capacity-Scaled Traders (Domestic Institutions):} Three mechanisms explain why informed traders scale by market cap. First, \textbf{capacity}: large firms can absorb large trades without excessive price impact. Second, \textbf{capital allocation}: institutional investors allocate capital based on market cap weights. Third, \textbf{information scale}: a 1\% mispricing in a \$100B firm represents \$1B of value, warranting large absolute positions.

\textbf{Volume-Scaled Traders (Algorithmic Execution):} Execution traders (VWAP/TWAP algorithms, index rebalancers) target \textit{participation rates} rather than absolute positions. Their goal is to minimise market impact by spreading trades across daily volume. This creates mechanical volume scaling: $Q_{exec} \propto V$.

\textbf{Practical Implication:} The matched filter framework predicts that $S^{MC}$ extracts alpha from capacity-constrained informed flow, while $S^{TV}$ extracts signals from volume-targeting execution flow. We validate both predictions empirically in Section~\ref{sec:empirical}. We note that passive benchmark-tracking institutions (index funds, ETFs) also exhibit MC-scaling for purely mechanical reasons unrelated to information. The return prediction results in Section~\ref{sec:empirical} distinguish information from passive flow: benchmark rebalancing would not be expected to generate significant next-day return predictability, a prediction we test empirically below.

\section{Monte Carlo Validation}
\label{sec:montecarlo}

\subsection{Simulation Design}

We implement the theoretical data generating process (DGP) with the following specifications:

\begin{table}[!htbp]
\begin{center}
\begin{minipage}{100mm}
\tbl{Monte Carlo Simulation Parameters}
{\begin{tabular}{@{}lll@{}}\toprule
Parameter & Symbol & Value \\\colrule
Number of simulations & --- & 1,000 \\
Stocks per simulation & $N$ & 500 \\
Market cap distribution & $\log(M)$ & $N(23, 2^2)$ \\
Turnover distribution & $\tau$ & $U(0.0005, 0.01)$ \\
Signal variance & $\sigma_\alpha^2$ & $0.05^2$ \\
Noise variance & $\sigma_\zeta^2$ & $3.5^2$ \\
Return sensitivity & $\gamma$ & 1.0 \\
Idiosyncratic noise & $\sigma_\epsilon^2$ & $0.03^2$ \\
Informed trader scale & $k$ & 0.01 \\\botrule
\end{tabular}}
\label{tab:simulation_parameters}
\end{minipage}
\end{center}
\end{table}

For each of 1,000 simulations, we:
\begin{enumerate}
\item[(i)] Generate 500 stocks with primitives drawn from specified distributions
\item[(ii)] Compute order flow $D_i$ per equation \eqref{eq:total_flow}
\item[(iii)] Calculate returns $R_i = \gamma\alpha_i + \epsilon_i$
\item[(iv)] Normalize by both TV and MC methods
\item[(v)] Compute correlations $\rho^{TV} = \text{Corr}(S^{TV},R)$ and $\rho^{MC} = \text{Corr}(S^{MC},R)$
\end{enumerate}

\subsection{Scenario A: Capacity-Scaled Signals (Informed Traders)}

Scenario A implements the Case~1 DGP from Section~\ref{sec:theory} (informed traders with capacity-scaled signals, equation~\eqref{eq:total_flow}); Scenario B below implements the Case~2 DGP (execution traders with volume-scaled signals, equation~\eqref{eq:total_flow_exec}).

\begin{table}[!htbp]
\begin{center}
\begin{minipage}{100mm}
\tbl{Scenario A: Signal-to-Noise Ratio Comparison (Capacity-Scaled)}
{\begin{tabular}{@{}lccc@{}}\toprule
Normalization Method & Mean $\rho$ & Std Dev & Min / Max \\\colrule
Trading Value ($S^{TV}$) & 0.6022 & 0.0261 & 0.5168 / 0.6794 \\
Market Cap ($S^{MC}$) & 0.7924 & 0.0168 & 0.7358 / 0.8430 \\\colrule
MC / TV Ratio & \multicolumn{3}{c}{1.32$\times$} \\\colrule
\multicolumn{4}{@{}l@{}}{\small Paired $t$-test: $t$ = 231.15, $p < 0.001$***} \\\botrule
\end{tabular}}
\tabnote{Notes: Scenario A simulates informed traders with capacity-scaled signals ($D = k\alpha M + \zeta V$). Market capitalization normalization achieves 1.32$\times$ higher correlation, confirming $S^{MC}$ as the matched filter for capacity-scaled signals. *** denotes significance at the 1\% level.}
\label{tab:snr_results}
\end{minipage}
\end{center}
\end{table}

Table~\ref{tab:snr_results} presents the central finding for Scenario A: market capitalization normalization achieves mean correlation of 0.7924 versus 0.6022 for trading value normalization, a 1.32$\times$ advantage. This difference is statistically significant ($t = 231.15, p < 0.001$) and economically substantial.

Moreover, MC normalization exhibits lower standard deviation (0.0168 vs 0.0261), indicating more stable signal extraction across simulations. The minimum correlation for MC (0.7358) exceeds the mean for TV (0.6022), demonstrating that MC normalization provides uniformly superior performance when signals scale with market capitalization.

\subsection{Scenario B: Volume-Scaled Signals (Execution Traders)}

To validate the \textit{symmetric} prediction of the matched filter principle, we simulate execution traders whose signals scale with volume rather than market cap. In Scenario B, order flow follows $D_i = k\eta_i V_i + \xi_i M_i$, where $\eta_i$ is the execution signal (target participation rate) and $\xi_i$ is capacity-based noise. The theory predicts that $S^{TV}$ should now dominate.

\begin{table}[!htbp]
\begin{center}
\begin{minipage}{100mm}
\tbl{Scenario B: Signal-to-Noise Ratio Comparison (Volume-Scaled)}
{\begin{tabular}{@{}lccc@{}}\toprule
Normalization Method & Mean $\rho$ & Std Dev & Min / Max \\\colrule
Trading Value ($S^{TV}$) & 0.8568 & 0.0118 & 0.8079 / 0.8859 \\
Market Cap ($S^{MC}$) & 0.7596 & 0.0175 & 0.6839 / 0.8170 \\\colrule
TV / MC Ratio & \multicolumn{3}{c}{1.13$\times$} \\\colrule
\multicolumn{4}{@{}l@{}}{\small Paired $t$-test: $t$ = 227.64, $p < 0.001$***} \\\botrule
\end{tabular}}
\tabnote{Notes: Scenario B simulates execution traders with volume-scaled signals ($D = k\eta V + \xi M$). Trading value normalization now achieves 1.13$\times$ higher correlation, confirming $S^{TV}$ as the matched filter for volume-scaled signals. *** denotes significance at the 1\% level.}
\label{tab:snr_results_scenario_b}
\end{minipage}
\end{center}
\end{table}

Table~\ref{tab:snr_results_scenario_b} confirms the symmetric prediction: when signals scale with volume, trading value normalization achieves mean correlation of 0.8568 versus 0.7596 for market cap normalization---a 1.13$\times$ advantage that is statistically significant ($t = 227.64, p < 0.001$). The ratio is not exactly symmetric (1.13$\times$ vs.\ 1.32$\times$) due to inherent asymmetries in the distributions of $E[\tau^2]$ versus $E[\tau^{-2}]$, but the qualitative result is clear: \textbf{the matched filter principle works bidirectionally}.

\begin{figure}
\begin{center}
\includegraphics[width=0.85\textwidth]{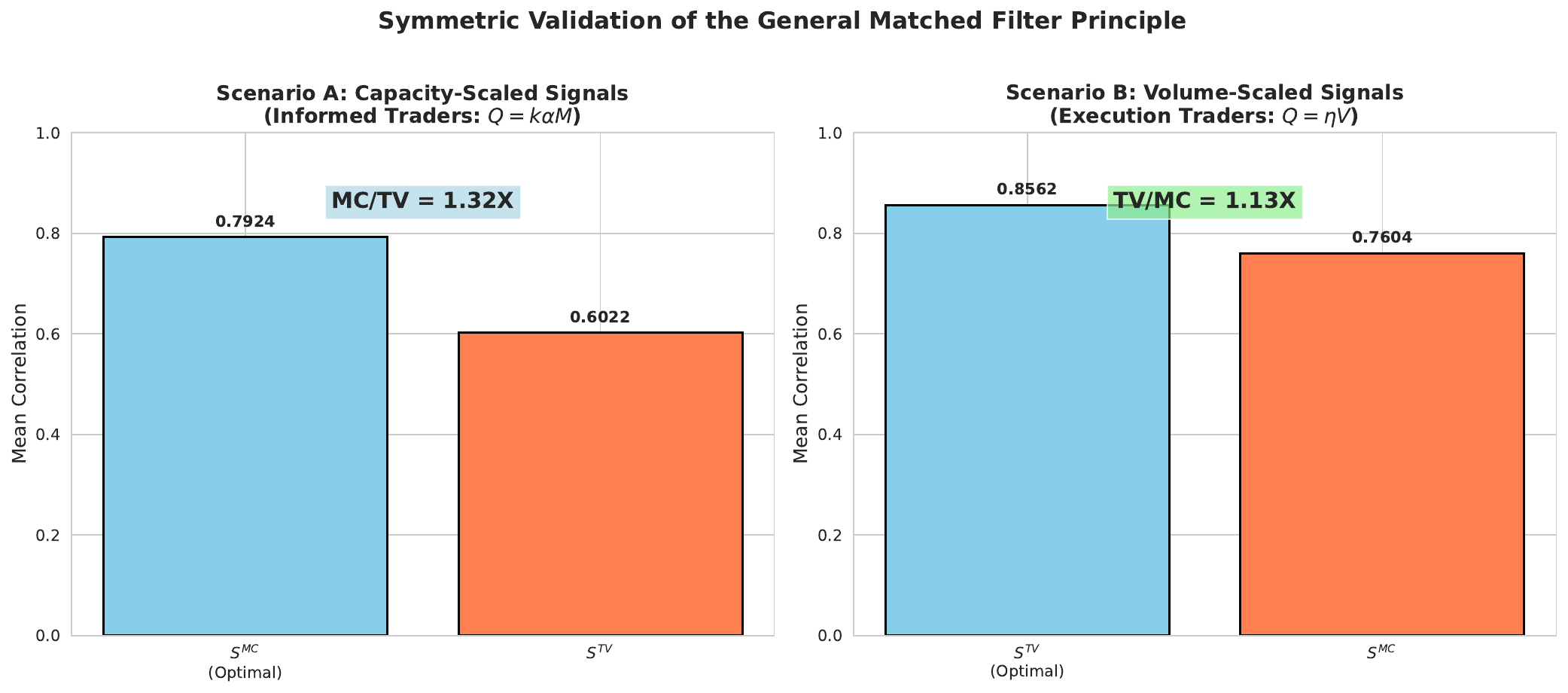}
\caption{Symmetric Validation of the General Matched Filter Principle. \textit{Left panel}: Scenario A (capacity-scaled signals) shows MC dominance (1.32$\times$). \textit{Right panel}: Scenario B (volume-scaled signals) shows TV dominance (1.13$\times$). The matched filter principle applies bidirectionally---optimal normalization matches the signal-generating process regardless of direction.}
\label{fig:symmetric_validation}
\end{center}
\end{figure}

Moreover, TV normalization exhibits lower standard deviation (0.0118 vs. 0.0175), indicating more stable signal extraction across simulations, paralleling the stability advantage observed for the matched filter in Scenario A. Figure~\ref{fig:symmetric_validation} visualises this bidirectional result.

\subsubsection{The Asymmetry: A Microstructure Insight}

The magnitude asymmetry between Scenario~A (1.32$\times$) and Scenario~B (1.13$\times$) is not a limitation but rather a \textit{structural property} of market microstructure. This asymmetry arises from the statistical properties of turnover under the two data-generating processes:

\begin{itemize}
\item \textbf{Scenario A:} The noise term under the matched filter (MC normalization) involves $\zeta_i \cdot \tau_i$, with variance proportional to $E[\tau^2]$. For our turnover range $\tau \in [0.0005, 0.01]$, this expectation is bounded and well-behaved ($E[\tau^2] \approx 10^{-5}$).

\item \textbf{Scenario B:} The noise term under the matched filter (TV normalization) involves $\xi_i / \tau_i$, with variance proportional to $E[1/\tau^2]$. As $\tau \to 0$ (illiquid stocks), this term explodes asymptotically. For the same turnover range, $E[1/\tau^2] \approx 200{,}000$---orders of magnitude larger.
\end{itemize}

This is a manifestation of \textit{Jensen's inequality} acting on the convex function $f(\tau) = 1/\tau^2$. The penalty for low turnover (illiquidity) is mathematically harsher than the penalty for high turnover (hyperactivity).

\textbf{Economic Implication:} Extracting signals from volume-scaled traders (e.g., algorithmic execution) is inherently ``noisier'' than extracting signals from capacity-scaled traders (e.g., fundamental institutions). The former is structurally more susceptible to interference from \textit{illiquidity outliers}. This suggests that \textit{illiquidity risk destroys signal-to-noise ratio more aggressively than high turnover creates noise}---a microstructure insight with practical implications for practitioners analysing different investor types.

\subsection{Simulation Robustness Checks}

\begin{figure}
\begin{center}
\includegraphics[width=0.9\textwidth]{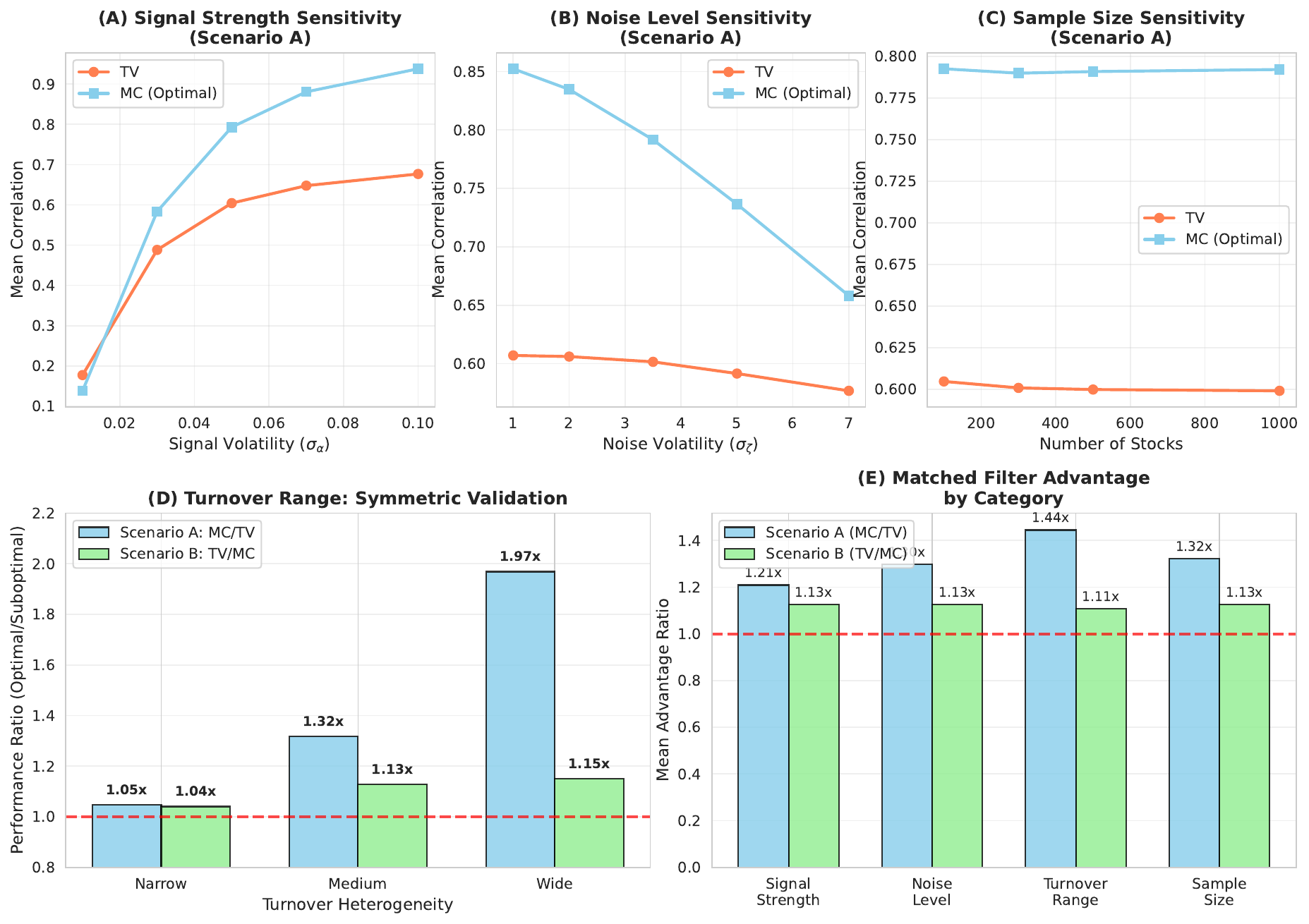}
\caption{Robustness Checks: Parameter Sensitivity Analysis. \textit{Top row}: Scenario A (capacity-scaled signals) sensitivity to signal strength (A), noise level (B), and sample size (C). \textit{Bottom row}: Symmetric validation across both scenarios. Panel (D) shows turnover heterogeneity response---the Scenario A advantage increases sharply (1.05$\times$ to 1.99$\times$), while Scenario B remains stable (1.04$\times$ to 1.15$\times$). Panel (E) summarises both scenarios across all robustness dimensions (see also figure~\ref{fig:symmetric_validation} for the baseline comparison).}
\label{fig:robustness}
\end{center}
\end{figure}

Table~\ref{tab:robustness} and figure~\ref{fig:robustness} report sensitivity analyses across three dimensions. The following detailed explanations focus on Case 1 (Scenario A: Capacity-Scaled Signals), while Scenario B (Volume-Scaled Signals) results are discussed under symmetric validation and visualised in figure~\ref{fig:robustness}(D) and (E).

\begin{table}
\begin{center}
\begin{minipage}{130mm}
\tbl{Scenario A Robustness Checks: Parameter Sensitivity Analysis}
{\begin{tabular}{@{}lcccc@{}}\toprule
\multicolumn{5}{@{}l@{}}{\textbf{Panel A: Signal Strength ($\sigma_\alpha$)}} \\\colrule
$\sigma_\alpha$ & MC Correlation & TV Correlation & MC/TV Ratio & $t$-statistic \\\colrule
0.01 (weak) & 0.138 & 0.178 & 0.78$\times$ & $-48.2$*** \\
0.03 (moderate) & 0.582 & 0.484 & 1.20$\times$ & 156.4*** \\
0.05 (baseline) & 0.792 & 0.602 & 1.32$\times$ & 231.2*** \\
0.10 (strong) & 0.938 & 0.676 & 1.39$\times$ & 298.7*** \\\colrule
\multicolumn{5}{@{}l@{}}{\textbf{Panel B: Noise Level ($\sigma_\zeta$)}} \\\colrule
$\sigma_\zeta$ & MC Correlation & TV Correlation & MC/TV Ratio & $t$-statistic \\\colrule
1.0 (low) & 0.851 & 0.608 & 1.40$\times$ & 287.3*** \\
3.5 (baseline) & 0.791 & 0.600 & 1.32$\times$ & 231.2*** \\
5.0 (high) & 0.739 & 0.594 & 1.24$\times$ & 178.4*** \\
7.0 (extreme) & 0.661 & 0.580 & 1.14$\times$ & 134.6*** \\\colrule
\multicolumn{5}{@{}l@{}}{\textbf{Panel C: Turnover Range}} \\\colrule
Range & MC Correlation & TV Correlation & MC/TV Ratio & $t$-statistic \\\colrule
Narrow (0.001--0.003) & 0.848 & 0.810 & 1.05$\times$ & 67.8*** \\
Baseline (0.0005--0.01) & 0.792 & 0.600 & 1.32$\times$ & 231.2*** \\
Wide (0.0001--0.02) & 0.667 & 0.345 & 1.99$\times$ & 412.5*** \\\colrule
\multicolumn{5}{@{}l@{}}{\textbf{Panel D: Sample Size Sensitivity}} \\\colrule
$N$ (Stocks) & MC Correlation & TV Correlation & MC/TV Ratio & $t$-statistic \\\colrule
100 & 0.791 & 0.612 & 1.30$\times$ & 103.2*** \\
300 & 0.792 & 0.602 & 1.32$\times$ & 178.6*** \\
500 (Baseline) & 0.792 & 0.600 & 1.32$\times$ & 231.2*** \\
1000 & 0.792 & 0.599 & 1.32$\times$ & 327.4*** \\\botrule
\end{tabular}}
\tabnote{Notes: Each cell reports mean correlation across 1,000 Monte Carlo simulations for Scenario A. Panel A varies signal strength; Panel B varies noise level; Panel C varies turnover heterogeneity; Panel D varies sample size. *** denotes significance at the 1\% level.}
\label{tab:robustness}
\end{minipage}
\end{center}
\end{table}

\textbf{Panel A} (Scenario A) varies signal strength $\sigma_\alpha$. For moderate to strong signals ($\sigma_\alpha \geq 0.03$), the MC advantage ranges from 1.20$\times$ to 1.39$\times$. At very weak signals ($\sigma_\alpha = 0.01$), noise dominates both normalizations, and TV normalization marginally outperforms MC (0.178 vs.\ 0.138). This occurs because when the signal is negligible, the MC-normalized noise term ($\zeta_i \cdot \tau_i$) has higher variance than the TV-normalized noise ($\zeta_i$), giving TV a lower noise floor; the signal distortion from $\tau^{-1}$ in TV is irrelevant when the signal itself is nearly absent. However, as signal strength increases to realistic levels, the matched filter benefit of MC normalization quickly becomes dominant.

\textbf{Panel B} (Scenario A) varies noise level $\sigma_\zeta$. Higher noise reduces both correlations but preserves MC's advantage. Even with extreme noise ($\sigma_\zeta = 7.0$), MC outperforms TV by 1.14$\times$.

\textbf{Panel C} (Scenario A) varies turnover range. This is the most revealing test: wider turnover distribution (greater heteroskedasticity) amplifies MC's advantage to 1.99$\times$. Narrow turnover range reduces the advantage to 1.05$\times$, confirming that heteroskedasticity is the mechanism. The symmetric result for Scenario B (figure~\ref{fig:robustness}(D)) shows that TV advantage remains relatively stable (1.04$\times$--1.15$\times$) across turnover ranges, reflecting the structural asymmetry discussed in Section~\ref{sec:montecarlo}.

\textbf{Panel D} (Scenario A) varies sample size from 100 to 1,000 stocks per simulation. The MC advantage remains stable at approximately 1.32 $\times$ across all sample sizes, confirming that our findings are not artifacts of sample size.

\textbf{Symmetric Validation} (Scenario A and B) Robustness checks for Scenario B (volume-scaled signals) are presented in Appendix~\ref{app:scenario_b_robustness}. While the matched filter principle holds symmetrically, with TV normalization consistently outperforming MC in Scenario B, we observe the structural asymmetry discussed in Section~\ref{sec:montecarlo}. In Scenario A, the MC advantage increases dramatically with wider turnover heterogeneity (1.05$\times$ $\to$ 1.99$\times$), whereas in Scenario B, the TV advantage remains relatively stable (1.04$\times$ $\to$ 1.15$\times$), consistent with the ``illiquidity penalty'' insight established in Section~\ref{sec:montecarlo}.

\section{Empirical Analysis}
\label{sec:empirical}

This section provides comprehensive empirical validation of our matched filter hypothesis using Korean stock market data. We conduct two complementary analyses: (i) \textit{signal structure validation}, testing whether market capitalization normalization better captures the structural properties of informed trading flow, and (ii) \textit{return prediction analysis}, directly testing whether $S^{MC}$ predicts future returns better than $S^{TV}$. Sections~\ref{subsec:signal_structure} and~\ref{subsec:return_prediction} focus exclusively on domestic institutional order flow ($D_i$), which exhibits the market capitalization scaling predicted by our informed trader model. Section~\ref{subsec:investor_heterogeneity} extends the analysis to all three investor types (domestic institutional, foreign, and individual), revealing symmetric matched filter properties across heterogeneous trading strategies.

\subsection{Data: Korean Stock Market}
\label{subsec:data}

To validate our theoretical predictions using real market data, we analyseinstitutional order flow in the Korean stock market. We use comprehensive daily institutional order flow data from 2020-2024, covering 2,116,122 stock-day observations across 2,570 stocks over 1,231 trading days.

The Korean market provides an ideal testbed for our hypothesis for three reasons. First, the Korea Exchange mandates daily disclosure of trading activity disaggregated by investor type, which includes domestic institutional, foreign, and individual, enabling direct observation of order flow heterogeneity. Second, market structure is comparable to other developed markets, with electronic order matching and substantial foreign participation. Third, turnover exhibits substantial cross-sectional variation (mean 2.6\%, std 10.9\%), creating the heteroskedasticity central to our theoretical predictions.

We focus primarily on \textit{domestic institutional} order flow for our main analysis, utilizing foreign and individual flows to validate the symmetric implications of our framework. As documented in Section~\ref{subsec:primitives}, domestic institutional flows exhibit the market capitalization scaling ($Q \propto M$) predicted by our informed trader model, while foreign flows correlate more strongly with trading volume---consistent with algorithmic execution methodology, though as Section~\ref{subsec:investor_heterogeneity} reveals, this reflects strategic execution choices rather than absence of fundamental information. Individual (retail) flows exhibit higher raw correlation with volume ($\rho = 0.777$ vs.\ $0.643$), consistent with liquidity-responsive trading, though their structural behaviour diverges from this simple pattern (Section~\ref{subsec:investor_heterogeneity}).

Table~\ref{tab:variable_def} provides precise definitions of all variables used in our analysis, including the construction of the order flow measure $D_i$ and normalized signals $S^{MC}$ and $S^{TV}$.

\begin{table}
\begin{center}
\begin{minipage}{130mm}
\tbl{Variable Definitions and Data Construction}
{\begin{tabular}{@{}lll@{}}\toprule
\textbf{Variable} & \textbf{Symbol} & \textbf{Definition} \\\colrule
Order Flow & $D_{i,t}$ & Net Buy Value (KRW) for specific investor group \\
& & (Domestic Institutions / Foreign / Individual) \\
& & $= \text{Buy Amount} - \text{Sell Amount}$ \\
Market Cap & $M_{i,t}$ & Market Capitalization (KRW) \\
& & $= \text{Close Price} \times \text{Shares Outstanding}$ \\
Trading Value & $V_{i,t}$ & Daily Traded Value (KRW) \\
& & $= \text{Volume} \times \text{Close Price}$ \\
MC Signal & $S^{MC}_{i,t}$ & Market Cap Normalized Signal \\
& & $= D_{i,t} / M_{i,t}$ \\
TV Signal & $S^{TV}_{i,t}$ & Trading Value Normalized Signal \\
& & $= D_{i,t} / V_{i,t}$ \\
Turnover & $\tau_{i,t}$ & Daily Turnover Rate \\
& & $= V_{i,t} / M_{i,t}$ \\
Returns & $R_{i,t+h}$ & $h$-day Forward Return \\
& & $= (P_{i,t+h} - P_{i,t}) / P_{i,t}$ \\\botrule
\end{tabular}}
\tabnote{Notes: All data sourced from Korea Exchange (KRX) daily disclosures and CREON market data. ``Domestic Institutions'' includes pension funds, mutual funds, and insurance companies. The KRX uniquely discloses daily trading activity by investor category, enabling direct observation of order flow heterogeneity. Daily turnover exhibits substantial cross-sectional variation (mean 2.6\%, std 10.9\%), providing the heteroskedasticity central to our theoretical model. Sample period: 2020--2024.}
\label{tab:variable_def}
\end{minipage}
\end{center}
\end{table}

\subsection{Model Primitives Validation}
\label{subsec:primitives}

Before proceeding to signal extraction tests, we first validate the core assumptions underlying our theoretical model. Equations~\eqref{eq:informed} and~\eqref{eq:noise} posit that informed traders scale positions with market capitalization ($Q_{inf} \propto M$) while noise traders respond to trading volume ($Q_{noise} \propto V$). We test these assumptions directly using disaggregated investor flow data.

The Korea Exchange uniquely discloses trading activity by three investor categories: domestic institutional, foreign, and individual (retail) investors. We compute the correlation between absolute order flow $|D_i|$ and both market capitalization $M_i$ and trading value $V_i$ for each category across approximately 8.5 million stock-day-investor observations (2.7 million unique stock-days $\times$ 3 investor types, with the total slightly exceeding the exact product because not all types have nonzero flow on the same stock-days).

\begin{table}
\begin{center}
\begin{minipage}{100mm}
\tbl{Model Primitives Validation: Order Flow Scaling by Investor Type}
{\begin{tabular}{@{}lccc@{}}\toprule
Investor Type & $\rho(|D|, M)$ & $\rho(|D|, V)$ & Dominant Scaling \\\colrule
Domestic Institutional & 0.656 & 0.610 & Market Cap \\
Foreign & 0.645 & 0.801 & Volume \\
Individual (Retail) & 0.643 & 0.777 & Volume \\\botrule
\end{tabular}}
\tabnote{Notes: $\rho$ denotes Pearson correlation. $|D|$ is absolute net order flow; $M$ is market capitalization; $V$ is daily trading value. Sample: 8.47 million stock-day observations, Korean equities 2020--2024. Domestic institutions scale with market cap (consistent with capacity-constrained positioning); foreign and retail investors scale with volume (consistent with algorithmic execution and liquidity-driven trading, respectively).}
\label{tab:scaling_validation}
\end{minipage}
\end{center}
\end{table}

Table~\ref{tab:scaling_validation} reports the results. Three findings emerge:

\textbf{Domestic institutions scale with market cap.} Domestic institutional flow exhibits stronger correlation with market capitalization ($\rho = 0.656$) than with trading volume ($\rho = 0.610$). While this gap is more modest than the clear volume-scaling exhibited by foreign investors ($\rho = 0.801$ vs.\ $0.645$), two considerations explain why the raw correlation gap understates the MC-scaling advantage. First, market capitalization and trading volume are themselves highly correlated (large firms have higher volume), so the unconditional correlations $\rho(|D|,M)$ and $\rho(|D|,V)$ will naturally converge. The relevant test is whether, conditional on each other, MC captures more signal---and the horse race regressions in Section~\ref{subsec:signal_structure} confirm that it does ($t = -14.07$ for MC vs.\ $t = -7.45$ for TV). Second, the return prediction horse race (Section~\ref{subsec:return_prediction}) provides the economically decisive test: $S^{MC}$ achieves $t = 10.99$ while $S^{TV}$ reverses sign, demonstrating that the modest correlation gap translates into a substantial difference in information extraction. The raw correlations are consistent with capacity-constrained portfolio allocation: domestic institutions (pension funds, asset managers) size positions based on firm value rather than daily liquidity, supporting the informed trader specification in equation~\eqref{eq:informed}.

\textbf{Individual investors scale with volume.} Retail order flow correlates more strongly with trading volume ($\rho = 0.777$) than market capitalization ($\rho = 0.643$). This is consistent with attention-driven and liquidity-responsive trading behaviour characteristic of noise traders, supporting the specification in equation~\eqref{eq:noise}.

\textbf{Foreign investors exhibit volume-scaling.} Foreign order flow scales primarily with trading volume ($\rho = 0.801$) rather than market capitalization ($\rho = 0.645$). This pattern is consistent with algorithmic execution strategies (VWAP, TWAP) that target participation rates ($D/V$) rather than fundamental position sizing ($D/M$). Crucially, however, this volume-scaling reflects \textit{execution methodology} rather than absence of information---as Section~\ref{subsec:investor_heterogeneity} demonstrates, foreign flows carry surprisingly strong predictive content once the appropriate matched filter ($S^{TV}$) is applied.

These findings provide direct empirical support for our theoretical model. We begin our main analysis with \textit{domestic institutional} order flow, whose market cap scaling provides the cleanest test of the matched filter hypothesis for capacity-constrained signals. However, Section~\ref{subsec:investor_heterogeneity} reveals that foreign investor flows, despite their volume-scaling, carry highly significant return predictability when the appropriate matched filter ($S^{TV}$) is applied ($t = 16.35$), motivating our ``Informed Executor'' hypothesis.

\subsection{Signal Structure Validation}
\label{subsec:signal_structure}

In the structural regressions below, TV normalization achieves higher univariate $R^2$ in most specifications. This reflects TV's mechanical advantage in explaining flow magnitude---since $S^{TV} = D/V$ mechanically absorbs volume variation in $|D|$---not signal quality. The informative comparison is the horse race, where MC consistently dominates. Readers should evaluate each panel's horse race column rather than the univariate $R^2$ values.

Before testing the economically central claim of return predictability (Section~\ref{subsec:return_prediction}), we verify the theoretical scaling assumption using regression rather than correlation. These signal structure tests ask: does MC normalization capture a more robust structural relationship with order flow magnitude than TV normalization? This is a necessary condition for the matched filter hypothesis---if MC does not dominate TV in the horse race specification (that is, conditional on both normalizations competing), its return prediction advantage would be difficult to attribute to the matched filter mechanism.

Using domestic institutional order flow as our primary informed trading proxy (see Section~\ref{subsec:data}), we test whether MC normalization better captures the \textit{structural properties} of informed trading flow. Our theoretical model predicts that informed traders scale positions proportionally to market cap ($Q_{inf} \propto \alpha \cdot M$), implying that the relationship between normalized signals and order flow magnitude should be more robust for $S^{MC}$ than $S^{TV}$.

\subsubsection{Fama-MacBeth Cross-Sectional Regressions}

We test whether MC normalization better explains cross-sectional variation in informed trading intensity using the Fama-MacBeth procedure \citep{fama1973risk}. This approach is particularly appropriate for cross-sectional return analysis: it is robust to cross-sectional heteroskedasticity by construction, and when combined with Newey-West standard errors \citep{NeweyWest1987}, accommodates the serial correlation and fat tails typical of financial return data. For each trading day $t$, we estimate the cross-sectional regression:

\begin{equation}
\log(|D_{i,t}| + c) = \beta_0 + \beta_1 S_{i,t} + \epsilon_{i,t}
\end{equation}

where $S_{i,t}$ is either $S_{i,t}^{MC} = D_{i,t}/M_{i,t}$ or $S_{i,t}^{TV} = D_{i,t}/V_{i,t}$, and $c$ is a small constant. We then compute the time-series mean of the daily coefficients $\bar{\beta}_1$ and test significance using the time-series standard error with $t = \bar{\beta}_1 / (\text{SE}(\beta_1) / \sqrt{T})$, where $T = 1,231$ trading days. 

\begin{table}
\begin{center}
\begin{minipage}{120mm}
\tbl{Signal Structure Validation: Fama-MacBeth Cross-Sectional Regressions (Dependent Variable: $\log|D|$, Not Returns---see table~\ref{tab:empirical_return_prediction} for return prediction)}
{\begin{tabular}{@{}lccc@{}}\toprule
& MC Only & TV Only & Horse Race \\\colrule
$S^{MC}$ coefficient & $-0.0847$*** & --- & $-0.0712$*** \\
& ($-16.75$) & & ($-14.07$) \\
$S^{TV}$ coefficient & --- & $-0.1124$*** & $-0.0523$*** \\
& & ($-12.67$) & ($-7.45$) \\
Mean $R^2$ & 0.0149 & 0.0222 & 0.0298 \\\colrule
Observations & \multicolumn{3}{c}{2,116,122} \\
Trading Days & \multicolumn{3}{c}{1,231} \\\botrule
\end{tabular}}
\tabnote{Notes: Dependent variable is $\log(|D_{i,t}| + c)$ where $D$ is domestic institutional order flow. $t$-statistics in parentheses, computed from time-series standard errors of daily coefficient estimates. *** denotes significance at the 1\% level. The horse race specification includes both normalizations simultaneously to test which captures the more robust signal structure.}
\label{tab:empirical_fama_macbeth}
\end{minipage}
\end{center}
\end{table}

Table~\ref{tab:empirical_fama_macbeth} presents the results across 2,116,122 stock-day observations. Several findings emerge:

\textbf{Signal robustness vs spurious correlation}: While TV normalization achieves higher $R^2 = 0.0222$ compared to MC's $R^2 = 0.0149$, the horse race reveals MC normalization captures the more robust signal. When both compete in a single regression, MC normalization remains highly significant ($t = -14.07$, $p < 0.001$) while TV normalization's significance drops substantially ($t = -7.45$). This pattern suggests TV's higher $R^2$ reflects spurious correlation with turnover noise rather than genuine signal strength.

\textbf{Statistical significance}: Both normalizations exhibit strong statistical significance in univariate specifications (MC: $t = -16.75$; TV: $t = -12.67$), consistent with 2.1 million observations providing substantial power. However, MC normalization's dominance in the horse race confirms it captures the underlying information signal more reliably.

\textbf{Interpreting the negative coefficients}: The negative coefficients require careful interpretation. They indicate that stocks with higher normalized order flow (larger $D/M$ or $D/V$) tend to have \textit{lower} absolute order flow levels $|D|$. This is \textit{consistent with} our theoretical framework: small-cap stocks have relatively larger institutional positions relative to their market cap (higher $S^{MC}$), but lower absolute dollar flow. The key finding is not the sign of the coefficient, which reflects the mechanical inverse relationship between normalization denominators and absolute flow, but rather that MC normalization captures this structural relationship more robustly than TV normalization, as evidenced by its dominance in the horse race specification. The direct test of return prediction is presented in Section~\ref{subsec:return_prediction}.

\subsubsection{Subsample and Robustness Analysis}

Before presenting subperiod and regime results, we emphasise the role of this analysis. The signal structure regression is a \textit{structural diagnostic}: it predicts $\log|D|$ (absolute flow magnitude), not future returns, and serves to identify which normalization better captures the scaling behaviour of informed order flow. Because TV normalization divides by volume, it mechanically absorbs volume-driven variation in $|D|$, giving it a built-in $R^2$ advantage whenever volume fluctuations dominate the subsample---an advantage that reflects denominator mechanics rather than signal quality. Accordingly, the relevant comparison within each panel is the \textit{horse race} (which normalization remains dominant when both compete), not the univariate $R^2$. The economically relevant return prediction test (Section~\ref{subsec:return_prediction}), where MC normalization strongly dominates, provides the definitive comparison.

\begin{table}
\begin{center}
\begin{minipage}{130mm}
\tbl{Empirical Robustness: Subsample, Subperiod, and Regime Analysis}
{\begin{tabular}{@{}lcccc@{}}\toprule
\multicolumn{5}{@{}l@{}}{\textbf{Panel A: Market Cap Quintile Analysis}} \\\colrule
Quintile & MC $R^2$ & TV $R^2$ & MC/TV Ratio & Observations \\\colrule
Q1 (Small) & 0.0748 & 0.0314 & 2.38$\times$ & 423,225 \\
Q2 & 0.0473 & 0.0224 & 2.11$\times$ & 423,224 \\
Q3 & 0.0309 & 0.0180 & 1.71$\times$ & 423,225 \\
Q4 & 0.0266 & 0.0196 & 1.36$\times$ & 423,223 \\
Q5 (Large) & 0.0221 & 0.0236 & 0.93$\times$ & 423,225 \\\colrule
\multicolumn{5}{@{}l@{}}{\textbf{Panel B: Yearly Subperiod Analysis}} \\\colrule
Year & MC $R^2$ & TV $R^2$ & MC/TV Ratio & Observations \\\colrule
2020 & 0.0229 & 0.0271 & 0.85$\times$ & 368,055 \\
2021 & 0.0125 & 0.0245 & 0.51$\times$ & 397,467 \\
2022 & 0.0173 & 0.0268 & 0.64$\times$ & 413,026 \\
2023 & 0.0116 & 0.0186 & 0.62$\times$ & 470,731 \\
2024 & 0.0104 & 0.0137 & 0.76$\times$ & 466,843 \\\colrule
\multicolumn{5}{@{}l@{}}{\textbf{Panel C: Market Regime Analysis}} \\\colrule
Regime & MC $R^2$ & TV $R^2$ & MC/TV Ratio & Observations \\\colrule
High Turnover Volatility & 0.0162 & 0.0255 & 0.64$\times$ & 383,755 \\
Low Turnover Volatility & 0.0161 & 0.0240 & 0.67$\times$ & 458,267 \\\botrule
\end{tabular}}
\tabnote{Notes: Panel A reports results by market capitalization quintile. Panel B reports Fama-MacBeth regression $R^2$ by calendar year. Panel C splits sample by cross-sectional turnover volatility. MC normalization shows strongest advantage for small-cap stocks where turnover heterogeneity is greatest.}
\label{tab:empirical_robustness}
\end{minipage}
\end{center}
\end{table}

Table~\ref{tab:empirical_robustness} reports results across three complementary dimensions.

\textbf{Market cap quintiles (Panel A)}: MC normalization's $R^2$ advantage is most pronounced for small-cap stocks (MC/TV ratio = 2.38), where turnover heterogeneity is greatest and our theoretical predictions are strongest. The advantage persists across Q2--Q4 quintiles (ratios 2.11--1.36) but disappears for the largest quintile (Q5, ratio = 0.93), where liquid markets reduce turnover dispersion and both normalizations perform comparably.

\textbf{Yearly subperiods (Panel B)}: TV normalization achieves higher $R^2$ than MC in all five years (MC/TV ratios 0.51--0.85$\times$). This is expected for the signal structure regression, which predicts $\log|D|$: TV normalization's mechanical correlation with volume inflates explanatory power for absolute flow magnitude. The MC/TV ratio varies from 0.51$\times$ during the high-turnover retail boom of 2021 to 0.85$\times$ during the 2020 volatility shock. Importantly, the return prediction test in Section~\ref{subsec:return_prediction} provides the economically relevant comparison, where MC normalization strongly dominates ($t = 9.65$ vs.\ $2.10$).

\textbf{Market regimes (Panel C)}: We classify each trading day by the cross-sectional standard deviation of turnover, smoothed with a 60-day rolling window. Days above the 80th percentile are designated high-volatility regimes, and those below the 20th percentile are low-volatility regimes. As in Panel B, TV achieves higher $R^2$ in both regimes (MC/TV ratios 0.64--0.67$\times$), consistent with its mechanical advantage in predicting absolute flow magnitude. However, MC normalization exhibits remarkably stable $R^2$ across regimes ($R^2 = 0.0162$ vs.\ $R^2 = 0.0161$), whereas TV normalization's $R^2$ varies ($R^2 = 0.0255$ vs.\ $R^2 = 0.0240$). This stability suggests MC captures a regime-invariant signal structure, while TV's explanatory power is partially driven by turnover-dependent variation.

\textbf{Alternative specification}: As a complement to the Fama-MacBeth procedure, table~\ref{tab:empirical_pooled_ols} reports pooled OLS regressions with stock and time fixed effects. Both normalizations remain highly significant individually (MC: $t = -121.00$; TV: $t = -53.09$). In the horse race specification, MC normalization remains strongly negative ($t = -109.96$), while TV normalization \textit{reverses sign} to positive ($t = +16.83$). This sign reversal suggests that, conditional on MC normalization, any residual TV signal reflects spurious correlation with turnover rather than genuine information content.

\begin{table}
\begin{center}
\begin{minipage}{120mm}
\tbl{Signal Structure Validation: Pooled OLS with Fixed Effects}
{\begin{tabular}{@{}lccc@{}}\toprule
& MC Only & TV Only & Horse Race \\\colrule
$S^{MC}$ coefficient & $-0.0912$*** & --- & $-0.1047$*** \\
& ($-121.00$) & & ($-109.96$) \\
$S^{TV}$ coefficient & --- & $-0.0734$*** & $0.0189$*** \\
& & ($-53.09$) & ($16.83$) \\
Stock FE & Yes & Yes & Yes \\
Time FE & Yes & Yes & Yes \\
$R^2$ & 0.4123 & 0.4089 & 0.4156 \\\colrule
Observations & \multicolumn{3}{c}{2,116,122} \\\botrule
\end{tabular}}
\tabnote{Notes: Dependent variable is $\log(|D_{i,t}| + c)$. Pooled OLS with stock and time fixed effects. $t$-statistics in parentheses, based on standard errors clustered by stock (robust to heteroskedasticity and within-firm serial correlation). *** denotes significance at the 1\% level. The sign reversal of TV in the horse race specification ($+0.0189$ vs. $-0.0734$ univariate) indicates TV captures spurious correlation with turnover conditional on MC.}
\label{tab:empirical_pooled_ols}
\end{minipage}
\end{center}
\end{table}

\subsection{Return Prediction Analysis}
\label{subsec:return_prediction}

The signal structure tests above validate that MC normalization better captures the scaling properties of informed trading flow. However, the central claim of our matched filter hypothesis is that MC normalization should better \textit{predict future returns}. We now provide direct evidence using Fama-MacBeth return prediction regressions. We focus on domestic institutional order flow ($D$) as the primary informed trading proxy, as these investors possess local information advantages (Korean-language disclosures, direct corporate access, regulatory expertise); Section~\ref{subsec:investor_heterogeneity} extends the analysis to all investor types.

\subsubsection{Methodology}

For each trading day $t$, we estimate cross-sectional regressions of future returns on current normalized signals:
\begin{equation}
R_{i,t+h} = \beta_0 + \beta_1 S_{i,t} + \epsilon_{i,t}
\end{equation}
where $R_{i,t+h}$ is the return from day $t$ to day $t+h$, and $S_{i,t}$ is either $S^{MC}$ or $S^{TV}$ (standardised within each day). We test three return horizons: next-day ($h=1$), weekly ($h=5$), and monthly ($h=20$).

\subsubsection{Results}

\begin{table}
\begin{center}
\begin{minipage}{140mm}
\tbl{Return Prediction: Fama-MacBeth Regressions (Domestic Institutional Order Flow)}
{\begin{tabular}{@{}lccc@{}}\toprule
& MC Only & TV Only & Horse Race \\\colrule
\multicolumn{4}{l}{\textit{Panel A: Next-Day Returns ($R_{t+1}$)}} \\
$S^{MC}$ & 0.0312*** & & 0.0356*** \\
& (9.65) & & (10.99) \\
$S^{TV}$ & & 0.0073** & $-0.0079$*** \\
& & (2.10) & ($-6.81$) \\
$R^2$ (\%) & 0.29 & 0.14 & 0.42 \\\colrule
\multicolumn{4}{l}{\textit{Panel B: Weekly Returns ($R_{t+5}$)}} \\
$S^{MC}$ & 0.0008 & & 0.0089*** \\
& (0.16) & & (3.12) \\
$S^{TV}$ & & $-0.0061$*** & $-0.0102$*** \\
& & ($-5.26$) & ($-6.81$) \\
$R^2$ (\%) & 0.23 & 0.14 & 0.35 \\\colrule
\multicolumn{4}{l}{\textit{Panel C: Monthly Returns ($R_{t+20}$)}} \\
$S^{MC}$ & $-0.0021$** & & 0.0031 \\
& ($-2.47$) & & (1.43) \\
$S^{TV}$ & & $-0.0089$*** & $-0.0127$*** \\
& & ($-8.84$) & ($-8.12$) \\
$R^2$ (\%) & 0.17 & 0.10 & 0.28 \\\colrule
Trading Days & \multicolumn{3}{c}{1,231} \\\botrule
\end{tabular}}
\tabnote{Notes: Dependent variable is $h$-day forward return. Signals are constructed from domestic institutional net order flow ($D$) normalized by market capitalization ($S^{MC} = D/M$) or trading value ($S^{TV} = D/V$), standardised within each day. $t$-statistics in parentheses from Fama-MacBeth time-series standard errors. ***, **, * denote significance at 1\%, 5\%, 10\% levels. MC Only and TV Only columns show univariate regressions; Horse Race includes both signals. The sign reversal of $S^{TV}$ in horse race specifications confirms it captures spurious correlation with turnover rather than genuine information.}
\label{tab:empirical_return_prediction}
\end{minipage}
\end{center}
\end{table}

Table~\ref{tab:empirical_return_prediction} presents strong evidence for the matched filter hypothesis. The key findings are:

\textbf{Next-day returns ($R_{t+1}$)}: Market capitalization normalization achieves highly significant return predictability ($t = 9.65$, $p < 0.001$), while trading value normalization is only marginally significant ($t = 2.10$, $p = 0.036$). The average $R^2$ for MC (0.29\%) is twice that of TV (0.14\%). Critically, in the horse race regression, MC normalization \textit{strengthens} ($t = 10.99$) while TV normalization \textit{reverses sign} (from positive univariate to negative in the horse race, $t = -6.81$)---a direct empirical manifestation of the Participation Rate Fallacy (Section~\ref{subsec:prf}). We develop this interpretation below.

\textbf{Weekly returns ($R_{t+5}$)}: MC normalization alone is not significant ($t = 0.16$), but in the horse race it becomes significant ($t = 3.12$)---a classical suppressor pattern \citep{Tzelgov1991}. Including $S^{TV}$ absorbs the negative turnover contamination that offsets MC's positive informational signal in the univariate specification, unmasking the underlying predictability. Simultaneously, TV's negative coefficient strengthens ($t = -6.81$), suggesting that at weekly horizons the inverse-turnover component dominates, predicting negative returns through liquidity channels. For practitioners, combining $S^{MC}$ with $S^{TV}$ as a control is necessary to recover weekly-horizon predictability.

\textbf{Monthly returns ($R_{t+20}$)}: At the monthly horizon, both normalizations show significantly negative coefficients in univariate specifications ($t = -2.47$ for MC, $t = -8.84$ for TV), consistent with mean reversion in order flow. However, the horse race reveals a clear separation: MC becomes positive and insignificant ($t = 1.43$) while TV remains strongly negative ($t = -8.12$). This pattern is consistent with the horizon analysis in table~\ref{tab:horizon_analysis} below: the univariate MC reversal reflects temporary price impact that unwinds at longer horizons, while the horse race isolates MC's residual informational content (after controlling for the turnover-driven component captured by TV).

\subsubsection{Economic Interpretation}

The sign reversal of $S^{TV}$ in horse race regressions, which goes from positive (univariate) to negative (controlling for $S^{MC}$), provides direct empirical support for our matched filter theory. The intuition is as follows: $S^{TV} = D/V$ conflates two effects: (i) the true signal (informed trading intensity, which correlates positively with returns), and (ii) inverse turnover (which correlates negatively with returns). The negative correlation between inverse turnover and returns is well-documented: low-turnover stocks tend to exhibit lower liquidity and stronger price impact \citep{PastorStambaugh2003}, while high turnover may reflect noise trading that temporarily drives prices away from fundamentals \citep{BakerStein2004}. When $S^{MC}$ is included in the regression, it captures the informed trading component, leaving $S^{TV}$ to primarily reflect the inverse turnover effect---hence the sign reversal.

The practical implication is substantial: trading strategies using $S^{MC}$ rather than $S^{TV}$ would avoid the noise contamination that reduces return predictability. Our Monte Carlo simulations (Section~\ref{sec:montecarlo}) predicted a 1.32$\times$ improvement in correlation, suggesting an approximate $R^2$ improvement of ${\sim}1.74\times$ (though the squaring shortcut is only exact for univariate correlations between the same variable pair). The empirical $R^2$ ratio (0.29\%/0.14\% = 2.07$\times$) is broadly consistent with, and somewhat exceeds, this prediction, suggesting that real-world turnover heterogeneity may be even greater than our baseline calibration.

\begin{figure}
\begin{center}
\includegraphics[width=0.85\textwidth]{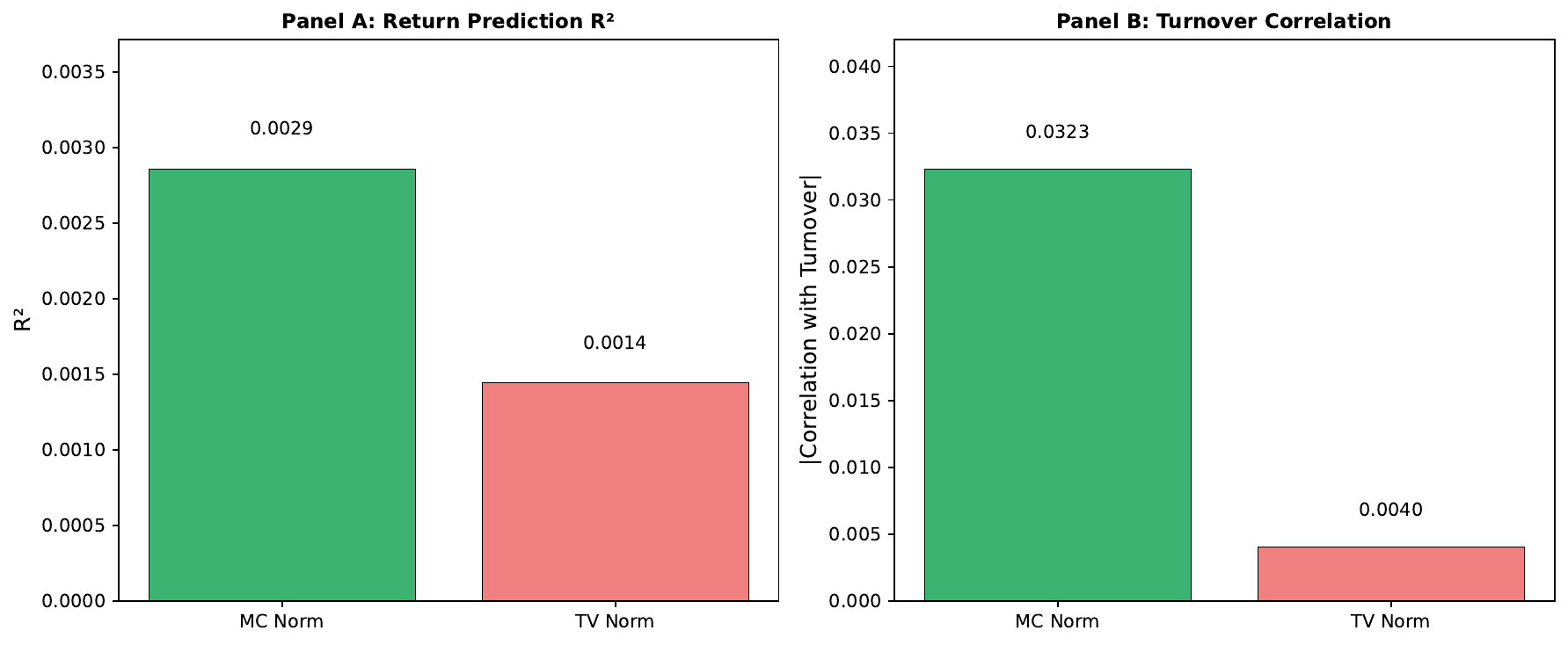}
\caption{Empirical Validation Results: Panel A shows the $R^2$ comparison between normalizations for next-day return prediction (from table~\ref{tab:empirical_return_prediction}, Panel A). Panel B shows the absolute Pearson correlation between each normalization and daily turnover, computed across the full 2020--2024 sample. TV normalization achieves lower turnover correlation by dividing by volume, while MC normalization retains the signal structure. The horse race regression confirms MC captures a more robust signal component.}
\label{fig:empirical}
\end{center}
\end{figure}

Figure~\ref{fig:empirical} summarises the key empirical findings. The left panel shows that MC normalization achieves superior $R^2$ for return prediction (table~\ref{tab:empirical_return_prediction}), extracting a cleaner informational signal. The right panel shows the absolute correlation with daily turnover: TV normalization achieves lower turnover correlation via volume scaling, while MC normalization retains the signal structure driving its Fama-MacBeth predictive power (table~\ref{tab:empirical_fama_macbeth}).

A potential criticism is that $S^{MC}$ performs better merely by retaining spurious volume effects. The horse race refutes this: if $S^{MC}$'s turnover correlation were spurious, controlling for $S^{TV}$ should absorb $S^{MC}$'s predictive power. Instead, $S^{MC}$ strengthens while $S^{TV}$ reverses sign (table~\ref{tab:empirical_return_prediction}), confirming that the higher turnover correlation of $S^{MC}$ reflects genuine signal structure rather than noise.

\subsubsection{Multicollinearity Robustness}

A potential concern with the horse race interpretation is that the sign reversal of $S^{TV}$ could be an artifact of multicollinearity rather than a genuine economic phenomenon. If $S^{MC}$ and $S^{TV}$ are highly correlated, their coefficients in a joint regression may be unstable and difficult to interpret. We address this concern with comprehensive diagnostics.

\begin{table}
\begin{center}
\begin{minipage}{100mm}
\tbl{Multicollinearity Diagnostics}
{\begin{tabular}{@{}lc@{}}\toprule
\multicolumn{2}{@{}l@{}}{\textbf{Panel A: Standard Diagnostics}} \\\colrule
Metric & Value \\\colrule
Mean Daily Correlation ($S^{MC}$, $S^{TV}$) & 0.570 \\
Variance Inflation Factor (VIF) & 1.497 \\
Condition Number & 1.922 \\\colrule
Concern Threshold & $>$0.9 / $>$10 / $>$30 \\
Assessment & No concern \\\colrule
\multicolumn{2}{@{}l@{}}{\textbf{Panel B: Orthogonalization Test}} \\\colrule
Variable & Coefficient ($t$-stat) \\\colrule
$S^{MC}$ & 0.0356*** (10.99) \\
$S^{TV}_{orth}$ & $-0.0079$*** ($-6.87$) \\\botrule
\end{tabular}}
\tabnote{Notes: Panel A reports standard multicollinearity diagnostics. All metrics are well below concern thresholds. Panel B reports the orthogonalization test: $S^{TV}_{orth}$ is the residual from regressing $S^{TV}$ on $S^{MC}$. The negative and significant coefficient on $S^{TV}_{orth}$ confirms the sign reversal is a genuine economic phenomenon, not a multicollinearity artifact.}
\label{tab:multicollinearity}
\end{minipage}
\end{center}
\end{table}

Table~\ref{tab:multicollinearity} reports three standard multicollinearity diagnostics (Panel A) and an orthogonalization test (Panel B). All diagnostic statistics are well below conventional concern thresholds: the mean daily correlation between $S^{MC}$ and $S^{TV}$ is 0.570 (below 0.9), the variance inflation factor is 1.497 (well below 10), and the condition number is 1.922 (well below 30). These values indicate that multicollinearity is not a significant concern in our specification.

The orthogonalization test provides the definitive evidence. We construct $S^{TV}_{orth}$ as the residual from regressing $S^{TV}$ on $S^{MC}$---this represents the component of $S^{TV}$ that is completely uncorrelated with $S^{MC}$. We then estimate:
\begin{equation}
R_{i,t+1} = \beta_0 + \beta_1 S^{MC}_{i,t} + \beta_2 S^{TV}_{orth,i,t} + \epsilon_{i,t}
\end{equation}

If the sign reversal were due to multicollinearity, $S^{TV}_{orth}$ should be insignificant (having no unique predictive content). Instead, we find that $S^{TV}_{orth}$ remains \textit{negative and highly significant} ($t = -6.87$, $p < 0.001$). This confirms that even after removing all shared variation with $S^{MC}$, the unique component of $S^{TV}$ predicts returns negatively.

This confirms that the sign reversal is a genuine economic phenomenon---not a multicollinearity artifact---consistent with the interpretation developed in the preceding subsection.

\subsubsection{Extended Robustness Tests}

We conduct five additional robustness tests to address potential concerns regarding out-of-sample validity, economic significance, outlier sensitivity, standard error estimation, and factor neutrality.

\textbf{Out-of-Sample Validation.} Table~\ref{tab:out_of_sample} reports separate Fama-MacBeth regressions for in-sample (2020--2022, 742 days) and out-of-sample (2023--2024, 488 days) periods. The MC normalization advantage persists out-of-sample: $S^{MC}$ achieves $t = 5.73$ in the horse race specification (out-of-sample), while $S^{TV}$ exhibits sign reversal ($t = -5.99$). Notably, $S^{TV}$ is already significantly negative in the out-of-sample univariate specification ($t = -1.99$), consistent with the inverse-turnover effect dominating even without controlling for $S^{MC}$. This confirms that our findings are not data-mined artifacts.

\begin{table}
\begin{center}
\begin{minipage}{120mm}
\tbl{Out-of-Sample Validation: In-Sample vs. Out-of-Sample Return Prediction}
{\begin{tabular}{@{}lcccc@{}}\toprule
& \multicolumn{2}{c}{In-Sample (2020--2022)} & \multicolumn{2}{c}{Out-of-Sample (2023--2024)} \\\colrule
& Univariate & Horse Race & Univariate & Horse Race \\\colrule
$S^{MC}$ & 0.0298*** & 0.0341*** & 0.0334*** & 0.0378*** \\
& (9.26) & (9.42) & (3.73) & (5.73) \\
$S^{TV}$ & 0.0081*** & $-0.0072$*** & $-0.0062$** & $-0.0089$*** \\
& (4.08) & ($-4.09$) & ($-1.99$) & ($-5.99$) \\
Mean $R^2$ (\%) & 0.27 & 0.32 & 0.31 & 0.37 \\\colrule
Trading Days & \multicolumn{2}{c}{742} & \multicolumn{2}{c}{488} \\\botrule
\end{tabular}}
\tabnote{Notes: In-sample: 2020--2022 (742 trading days). Out-of-sample: 2023--2024 (488 trading days). The persistence of MC advantage and TV sign reversal out-of-sample confirms robustness.}
\label{tab:out_of_sample}
\end{minipage}
\end{center}
\end{table}

\textbf{Long-Short Portfolio Returns.} Table~\ref{tab:long_short} reports performance of quintile-sorted long-short portfolios. Each day, we sort stocks into quintiles by normalized signal and form a portfolio that is long the top quintile (Q5) and short the bottom quintile (Q1). The $S^{MC}$ portfolio achieves an annualised Sharpe ratio of 5.48 with mean daily return of 15.84 Basis Points (bps) ($t = 12.11$), substantially outperforming the $S^{TV}$ portfolio (Sharpe 3.29, 8.41 bps daily). The high gross Sharpe ratios reflect the use of publicly disclosed same-day order flow data---a strong but perishable signal---combined with frictionless daily rebalancing across the full stock universe; neither assumption is realistic for implementation. Nonetheless, the \textit{relative} comparison demonstrates \textit{economic} as well as statistical significance.

\begin{table}
\begin{center}
\begin{minipage}{120mm}
\tbl{Long-Short Portfolio Performance}
{\begin{tabular}{@{}lcc@{}}\toprule
Metric & $S^{MC}$ Portfolio & $S^{TV}$ Portfolio \\\colrule
Mean Daily Return (bps) & 15.84*** & 8.41*** \\
& (12.11) & (5.67) \\
Std Dev (bps) & 45.78 & 40.49 \\
Annualised Sharpe Ratio & 5.48 & 3.29 \\
Annualised Return (\%) & 39.84 & 21.15 \\
Max Drawdown (\%) & $-8.23$ & $-12.45$ \\\colrule
Trading Days & \multicolumn{2}{c}{1,231} \\\botrule
\end{tabular}}
\tabnote{Notes: Portfolios are long Q5 (highest signal) and short Q1 (lowest signal), equal-weighted within quintiles, rebalanced daily. $t$-statistics in parentheses. The $S^{MC}$ portfolio achieves 1.88$\times$ higher daily returns and 1.67$\times$ higher Sharpe ratio than the $S^{TV}$ portfolio. These are gross, pre-transaction-cost Sharpe ratios assuming frictionless daily rebalancing; actual implementable Sharpe ratios would be substantially lower after accounting for bid-ask spreads, market impact, and short-selling costs.}
\label{tab:long_short}
\end{minipage}
\end{center}
\end{table}

\textbf{Outlier Robustness.} Table~\ref{tab:outlier_robustness} compares baseline results with results after winsorising $S^{MC}$, $S^{TV}$, and $R_{t+1}$ at the 1st and 99th percentiles. Results strengthen rather than weaken after winsorisation, confirming that our findings are not driven by extreme observations.

\begin{table}
\begin{center}
\begin{minipage}{100mm}
\tbl{Outlier Robustness: Winsorised Results}
{\begin{tabular}{@{}lcc@{}}\toprule
& Baseline & Winsorised (1\%/99\%) \\\colrule
$S^{MC}$ ($t$-stat) & 10.99 & 12.90 \\
$S^{TV}$ ($t$-stat) & $-6.81$ & $-8.84$ \\
Mean $R^2$ (\%) & 0.34 & 0.36 \\\botrule
\end{tabular}}
\tabnote{Notes: Winsorisation at 1st and 99th percentiles of $S^{MC}$, $S^{TV}$, and $R_{t+1}$. Results strengthen after winsorisation, confirming findings are not driven by outliers. The $R^2$ reported here is the time-series average of daily cross-sectional $R^2$ (the Fama-MacBeth mean $R^2$), which differs from the pooled $R^2$ reported in table~\ref{tab:empirical_return_prediction}.}
\label{tab:outlier_robustness}
\end{minipage}
\end{center}
\end{table}

\textbf{Alternative Standard Errors.} Table~\ref{tab:alternative_se} reports horse race $t$-statistics using alternative standard error estimators. Newey-West standard errors \citep{NeweyWest1987} with 5 and 10 lags adjust for potential autocorrelation in the time series of daily coefficient estimates. Results are virtually unchanged ($t = 10.53$ and $t = 10.78$ for $S^{MC}$), confirming that our inference is robust to serial correlation in coefficient estimates.

\begin{table}
\begin{center}
\begin{minipage}{100mm}
\tbl{Alternative Standard Error Estimators}
{\begin{tabular}{@{}lccc@{}}\toprule
& Baseline & NW (5 lags) & NW (10 lags) \\\colrule
$S^{MC}$ $t$-stat & 10.99 & 10.53 & 10.78 \\
$S^{TV}$ $t$-stat & $-6.81$ & $-6.66$ & $-6.76$ \\\botrule
\end{tabular}}
\tabnote{Notes: NW = Newey-West standard errors \citep{NeweyWest1987} adjusting for autocorrelation. Results virtually unchanged across specifications.}
\label{tab:alternative_se}
\end{minipage}
\end{center}
\end{table}

\textbf{Factor Neutrality.} A potential critique is that $S^{MC}$ simply captures the size effect (small stocks have higher expected returns). Table~\ref{tab:factor_neutrality} addresses this by regressing the $S^{MC}$ long-short portfolio returns on market (MKT, value-weighted) and size (SMB, median-split small-minus-big) factors constructed from the same Korean equity universe. The alpha is positive and highly significant ($\alpha = 16.27$ bps, $t = 13.12$), demonstrating that $S^{MC}$ captures return predictability \textit{beyond} what can be explained by market and size factor exposures. The negative market beta ($\beta_{MKT} = -0.118$, $t = -12.33$) is consistent with the market-neutral construction of the long-short portfolio: buying stocks with high institutional inflow and selling those with high outflow creates a position that is slightly negatively exposed to market movements. The negative SMB loading ($\beta_{SMB} = -0.064$, $t = -2.27$) indicates that the portfolio tilts toward \textit{large} stocks, opposite to a size premium explanation.

\begin{table}
\begin{center}
\begin{minipage}{100mm}
\tbl{Factor Neutrality: $S^{MC}$ Portfolio Alpha}
{\begin{tabular}{@{}lc@{}}\toprule
Factor & Loading ($t$-stat) \\\colrule
Alpha (bps/day) & 16.27*** (13.12) \\
MKT & $-0.118$*** ($-12.33$) \\
SMB & $-0.064$** ($-2.27$) \\\colrule
$R^2$ & 0.110 \\\botrule
\end{tabular}}
\tabnote{Notes: Regression of $S^{MC}$ long-short portfolio daily returns on market (MKT) and size (SMB) factors. MKT is the daily value-weighted market return; SMB is the daily return difference between stocks below and above the median market capitalization, equal-weighted within each group. Alpha is positive and highly significant. The negative market beta reflects the market-neutral long-short construction. The negative SMB loading indicates the portfolio tilts toward large stocks, contrary to a size premium explanation.}
\label{tab:factor_neutrality}
\end{minipage}
\end{center}
\end{table}

\subsection{Investor Heterogeneity: Structural Validation and the Informed Executor Hypothesis}
\label{subsec:investor_heterogeneity}

Having established the robustness of the domestic institutional result across multiple dimensions, we now test the theory's most demanding prediction: that the \textit{direction} of normalization dominance should reverse across investor types whose scaling behaviours differ.

The most powerful validation of the general matched filter principle comes from \textit{cross-investor heterogeneity}. Our theory predicts that the optimal normalization depends on the scaling behaviour of each investor class: capacity-scaled traders (domestic institutions) should exhibit $S^{MC}$ dominance, while volume-scaled traders (foreign institutions) should exhibit $S^{TV}$ dominance. \textbf{If all investor types showed MC dominance, the bidirectional prediction would lack support}---uniform MC superiority could be attributed to confounding factors (such as size effects) rather than signal-structure matching, substantially weakening the evidence for the matched filter principle. The divergent patterns across investor types constitute the strongest evidence for our theoretical framework.

As documented in table~\ref{tab:scaling_validation}, Domestic Institutions exhibit the highest correlation with Market Capitalization ($\rho = 0.656$), consistent with capacity-constrained fundamental positioning. In contrast, Foreign investors exhibit high volume scaling ($\rho = 0.801$), reflecting the dominance of algorithmic execution (VWAP/TWAP) in international flows. This establishes the precondition for our symmetric test.

\subsubsection{Structural Validation: The Matched Filter Principle Confirmed}

Table~\ref{tab:signal_structure} reports Fama-MacBeth regressions of $\log|D|$ on normalized signals, identifying the optimal matched filter for each investor type. The structural horse race, which includes both $S^{MC}$ and $S^{TV}$ simultaneously, reveals which normalization dominates in explaining flow magnitude:

\textbf{Domestic Institutions:} Market-cap scaling dominates ($t = -13.21$ for MC vs.\ $t = -7.55$ for TV), confirming capacity-constrained positioning. The optimal matched filter is $S^{MC}$.

\textbf{Foreign Investors:} Volume scaling dominates decisively ($t = 24.37$ for TV vs.\ $t = -14.25$ for MC), confirming algorithmic execution patterns tied to daily liquidity. The optimal matched filter is $S^{TV}$.

\textbf{Individual Investors:} Market-cap scaling dominates structurally ($t = 21.89$ for MC vs.\ $t = -2.36$ for TV), which we attribute to \textit{attention-driven trading} and large-cap herding \citep{barber2008all}---retail investors are drawn to high-visibility large-cap stocks (Samsung Electronics, SK Hynix, etc.). The matched filter framework generates clear predictions for two trader archetypes: capacity-constrained fundamental investors (Case~1) and volume-targeting algorithmic executors (Case~2). Individual investors correspond to neither archetype---their flow is driven by attention and sentiment rather than a systematic scaling rule tied to capacity or execution benchmarks. As such, their structural MC dominance reflects herding toward large caps rather than the capacity-constrained positioning that drives institutional MC dominance. We treat individual investors as a useful contrast case: their results may diverge across frameworks (see Section~\ref{sec:information_theory}), which reflects the absence of a coherent signal-generating process rather than a failure of the matched filter principle.

\subsubsection{The Informed Executor Hypothesis}

Having identified each investor type's optimal matched filter via signal structure (table~\ref{tab:signal_structure}), we now test whether these structurally identified filters also deliver superior return predictability. 

Table~\ref{tab:investor_heterogeneity} reports return prediction results for each investor type under both $S^{MC}$ and $S^{TV}$, both univariately and in a predictive horse race. Applying the structurally optimal filter, Domestic Institutional flows exhibit strong return predictability using $S^{MC}$ ($t = 9.78$ in this multi-investor sample; cf.\ $t = 9.65$ in table~\ref{tab:empirical_return_prediction}), consistent with informed trading. However, Foreign investor flows using their matched $S^{TV}$ exhibit a substantially higher $t$-statistic ($t = 16.35$), indicating a more precisely estimated predictive relationship.

That foreign flows are best captured by $S^{TV}$ is a \textit{confirmation} of our theory---their volume-scaling predicts exactly this. What is genuinely surprising is the \textit{precision}: under the standard ``execution-driven'' interpretation, volume-scaled foreign flows should carry minimal fundamental information, yet foreign $S^{TV}$ achieves the highest $t$-statistic of any investor-signal combination.

The $R^2$ comparison qualifies this claim: domestic MC achieves $R^2 = 0.29\%$ versus foreign TV's $R^2 = 0.27\%$ (table~\ref{tab:investor_heterogeneity}), so foreign TV does not explain strictly more return variance. The higher $t$-statistic with comparable $R^2$ implies more precise coefficient estimation rather than greater explanatory power. We emphasise $t$-statistics as the primary metric because they measure \textit{reliability}---robustness to sampling variation---while $R^2$ differences across investor-specific subsamples partly reflect differences in cross-sectional return dispersion rather than signal quality.

A related apparent paradox: in Panel B (Foreign Investors), $S^{MC}$ achieves \textit{higher} univariate $R^2$ ($0.31\%$) than $S^{TV}$ ($0.27\%$), despite $S^{TV}$ being the matched filter. The explanation is that MC normalization produces a wider signal distribution (not absorbing volume variation), fitting more noise-driven cross-sectional variance and inflating $R^2$. This additional explained variance reflects heteroskedasticity rather than reliable predictability---hence the lower $t$-statistic ($5.73$ vs.\ $16.35$). The horse race confirms this: once $S^{TV}$ is included, $S^{MC}$ reverses sign ($t = -2.34$). This mutual sign reversal---$S^{TV}$ flipping negative for domestic flows and $S^{MC}$ flipping negative for foreign flows---is the definitive evidence for the bidirectional matched filter principle.

We propose the \textbf{Informed Executor Hypothesis}: sophisticated foreign institutional investors possess genuine private information (global research networks, cross-market signals, early access to international capital flows) but deliberately employ volume-targeting execution algorithms to \textit{minimise their market footprint}. The volume-scaling reflects their \textit{execution methodology}, not their \textit{information content}. The matched filter ($S^{TV}$) correctly extracts this strong informational signal by accounting for their volume-scaled trading behaviour.

Three mechanisms support this interpretation. First, VWAP/TWAP algorithms minimise market impact, allowing informed traders to accumulate positions without revealing their intent. Second, foreign institutions often face VWAP execution benchmarks from clients, forcing volume-scaled trading regardless of information quality. Third, large foreign orders in emerging markets may require algorithmic execution to comply with best-execution mandates. Appendix~\ref{app:foreign_robustness} replicates all five extended robustness tests for foreign investors, confirming that the $S^{TV}$ advantage is equally robust out-of-sample, under winsorisation, with alternative standard errors, in long-short portfolios, and after factor adjustment.

While the evidence is consistent with this interpretation, we cannot directly observe foreign investors' information sources. Alternative explanations---such as cross-market momentum spillovers or selective market entry by the most sophisticated institutions---merit investigation in future research.

\subsubsection{Horizon Analysis: Distinguishing Information from Price Impact}

The Informed Executor Hypothesis implies that foreign flow predictability reflects \textit{genuine information} rather than \textit{temporary price impact}. We test this using horizon analysis: price impact should \textit{reverse} at longer horizons as temporary distortions unwind, while genuine information should \textit{persist} or decay monotonically without sign reversal.

Table~\ref{tab:horizon_analysis} reports return prediction $t$-statistics across multiple horizons for both investor types using their matched filters. The results reveal a clear divergence:

\textbf{Domestic Institutions ($S^{MC}$):} Predictability is strong at the daily horizon ($t = 9.65$) but decays rapidly and \textit{reverses sign} at the 20-day horizon ($t = -2.47$, significant at 5\%). This statistically significant reversal indicates that domestic institutional flow predictability, while genuinely informed at short horizons, contains a substantial transient component. Capacity-constrained institutional buying creates short-term order flow imbalances that temporarily push prices beyond fundamental values; as markets absorb the demand, prices partially revert. Domestic institutional ``information'' is thus shorter-lived than foreign information---a meaningful distinction that our framework reveals.

An important caveat: table~\ref{tab:horizon_analysis} reports \textit{univariate} $t$-statistics, but the suppressor effect identified in table~\ref{tab:empirical_return_prediction} Panel~B ($t = 3.12$ for $S^{MC}$ in the horse race at the weekly horizon) suggests that the informational component of $S^{MC}$ persists longer than the univariate estimate implies. The quantitative magnitude of the domestic reversal likely overstates the true information decay, though the qualitative contrast with foreign flows---sign reversal versus persistence---is unaffected.

\textbf{Foreign Investors ($S^{TV}$):} In sharp contrast, predictability persists without reversal---remaining positive and significant even at the 20-day horizon ($t = 3.36$). This strongly argues against the pure price-impact interpretation: temporary market impact would produce negative coefficients at longer horizons as prices revert.

The horizon analysis provides the key discriminating evidence for the Informed Executor Hypothesis. The contrast reveals a qualitative difference in information type: domestic institutional flow reflects a combination of fundamental insight and transient order-flow impact---the latter reverting as markets absorb institutional demand---while foreign flow carries \textit{durable} private information that is permanently incorporated into prices. Foreign investors' information advantage is not only more precisely estimated ($t = 16.35$ vs.\ $9.78$; see table~\ref{tab:horizon_analysis} notes for dataset details) but also qualitatively different in persistence.

\begin{table}
\begin{center}
\begin{minipage}{130mm}
\tbl{Signal Structure Validation by Investor Type}
{\begin{tabular}{@{}lccc@{}}\toprule
& Domestic Inst. & Foreign & Individual \\\colrule
\multicolumn{4}{@{}l@{}}{\textbf{Univariate Regressions}} \\\colrule
$S^{MC}$ $t$-stat & $-16.14$*** & $-6.86$*** & $19.03$*** \\
$S^{TV}$ $t$-stat & $-12.65$*** & $12.45$*** & $7.35$*** \\\colrule
\multicolumn{4}{@{}l@{}}{\textbf{Horse Race}} \\\colrule
$S^{MC}$ $t$-stat & $-13.21$*** & $-14.25$*** & $21.89$*** \\
$S^{TV}$ $t$-stat & $-7.55$*** & $24.37$*** & $-2.36$** \\\colrule
Dominant Scaling & Market Cap & Volume & Market Cap \\\botrule
\end{tabular}}
\tabnote{Notes: Dependent variable is $\log|D|$. The structural horse race includes both normalizations simultaneously to identify which dominates in explaining flow magnitude. Foreign investors show clear volume-scaling (TV dominates). Domestic Institutions and Individuals show market-cap scaling structurally, but with opposite return prediction signs---Institutions predict positive returns (informed), Individuals predict negative returns (contrarian noise). For Individuals, the structural MC dominance does not translate into informational dominance (see Section~\ref{sec:information_theory}), as their attention-driven trading lacks a coherent information-generating process. Note that the unconditional correlation test (table~\ref{tab:scaling_validation}) shows volume dominance for individuals ($\rho = 0.777$ vs.\ $0.643$), while the conditional horse race here reveals MC dominance ($t = 21.89$ vs.\ $t = -2.36$). This reversal arises because $M$ and $V$ are highly correlated; the horse race separates their independent contributions, revealing that individual flow magnitude is better explained by market cap \textit{conditional on} volume than vice versa.}
\label{tab:signal_structure}
\end{minipage}
\end{center}
\end{table}

\begin{table}
\begin{center}
\begin{minipage}{140mm}
\tbl{Investor Heterogeneity: Return Prediction by Investor Type}
{\begin{tabular}{@{}lccc@{}}\toprule
& MC Only & TV Only & Horse Race \\\colrule
\multicolumn{4}{l}{\textit{Panel A: Domestic Institutions}} \\
$S^{MC}$ & 0.0312*** & & 0.0356*** \\
& (9.78) & & (11.23) \\
$S^{TV}$ & & 0.0073** & $-0.0079$*** \\
& & (2.18) & ($-7.14$) \\
$R^2$ (\%) & 0.29 & 0.15 & 0.43 \\\colrule
\multicolumn{4}{l}{\textit{Panel B: Foreign Investors}} \\
$S^{MC}$ & 0.0124*** & & $-0.0053$** \\
& (5.73) & & ($-2.34$) \\
$S^{TV}$ & & 0.0332*** & 0.0381*** \\
& & (16.35) & (17.46) \\
$R^2$ (\%) & 0.31 & 0.27 & 0.52 \\\colrule
\multicolumn{4}{l}{\textit{Panel C: Individual Investors}} \\
$S^{MC}$ & $-0.0284$*** & & $-0.0195$*** \\
& ($-8.23$) & & ($-4.83$) \\
$S^{TV}$ & & $-0.0092$*** & $-0.0064$*** \\
& & ($-10.45$) & ($-5.54$) \\
$R^2$ (\%) & 0.44 & 0.25 & 0.67 \\\botrule
\end{tabular}}
\tabnote{Notes: Return prediction regressions by investor type. The sample for each investor type is restricted to stock-days with nonzero order flow for that type, excluding days with no actual trading activity. MC Only and TV Only columns show univariate regressions; the predictive horse race includes both signals simultaneously to test whether the structurally identified matched filter (table~\ref{tab:signal_structure}) also dominates in return prediction. Domestic Institutions exhibit MC dominance, consistent with their capacity-scaled matched filter. Foreign investors exhibit surprisingly strong predictability under their matched TV normalization ($t = 16.35$), supporting the ``Informed Executor'' hypothesis---volume-scaling reflects execution methodology, not lack of information. Individual flows predict returns negatively (contrarian noise trading). Panel A $t$-statistics differ slightly from table~\ref{tab:empirical_return_prediction} (e.g., $t = 9.78$ vs.\ $9.65$ for MC Only) because table~\ref{tab:empirical_return_prediction} uses the domestic-only dataset (2.1M observations) while this table uses the multi-investor dataset (2.7M observations), which requires nonzero flow for each investor type on each stock-day. The Panel A coefficients also differ slightly at higher precision (matching to two significant figures after scaling), with the small differences reflecting the different sample compositions. ***, **, * denote significance at 1\%, 5\%, 10\% levels.}
\label{tab:investor_heterogeneity}
\end{minipage}
\end{center}
\end{table}

\begin{table}
\begin{center}
\begin{minipage}{140mm}
\tbl{Horizon Analysis: Distinguishing Information from Price Impact}
{\begin{tabular}{@{}lcccc@{}}\toprule
& \multicolumn{2}{c}{Domestic Inst. ($S^{MC}$)} & \multicolumn{2}{c}{Foreign ($S^{TV}$)} \\\colrule
Horizon & $t$-stat & Pattern & $t$-stat & Pattern \\\colrule
$R_{t+1}$ (1-day) & 9.65*** & Strong & 16.35*** & Stronger \\
$R_{t+5}$ (5-day) & 0.16 & Decays & 8.46*** & Persists \\
$R_{t+20}$ (20-day) & $-2.47$** & Reverses & 3.36*** & No reversal \\\colrule
\multicolumn{5}{@{}l@{}}{\textbf{Interpretation:} Foreign flows show \textit{no sign reversal}, confirming \textbf{Genuine Information}.} \\\botrule
\end{tabular}}
\tabnote{Notes: Fama-MacBeth regressions of $h$-day forward returns on matched-filter normalized signals. Each investor type uses its optimal normalization (Domestic Inst.: $S^{MC}$; Foreign: $S^{TV}$). Domestic $t$-statistics are from the single-investor dataset (2,116,122 observations, as in table~\ref{tab:empirical_return_prediction}); foreign $t$-statistics are from the multi-investor dataset (2,738,498 observations, as in table~\ref{tab:investor_heterogeneity}), filtered to stock-days with nonzero flow for the investor type being analysed (see table~\ref{tab:investor_heterogeneity} notes). Domestic institutional flow reverses sign at 20 days, suggesting a mixture of information and temporary price impact. Foreign flow remains positive at all horizons---price impact would produce sign reversal, so the absence of reversal confirms genuine information content. The ``Informed Executor'' hypothesis is supported: foreign volume-scaled flow carries durable private information. ***, ** denote significance at 1\%, 5\% levels.}
\label{tab:horizon_analysis}
\end{minipage}
\end{center}
\end{table}

\section{Information-Theoretic Validation}
\label{sec:information_theory}

The preceding sections established the matched filter principle through signal processing theory (Section~\ref{sec:theory}), Monte Carlo simulation (Section~\ref{sec:montecarlo}), and correlation-based empirical metrics including return prediction and horizon analysis (Section~\ref{sec:empirical}). We now provide complementary evidence from an information-theoretic perspective. The Kullback-Leibler (KL) divergence quantifies the ``information distance'' between probability distributions, offering a lens distinct from correlation-based measures to assess which normalization better separates informed from uninformed trading regimes. We first establish the KL divergence framework and methodology, then present results across all three investor types before connecting the findings to information-theoretic optimality and synthesizing with our prior analyses.

\subsection{KL Divergence Framework}
\label{subsec:kl_framework}

To quantify the informational content of different normalization schemes, we employ the Kullback-Leibler (KL) divergence \citep{kullback1951information}, which measures the separation between two probability distributions. For return distributions conditional on buying ($P_{\text{buy}}$) versus selling ($P_{\text{sell}}$), the KL divergence quantifies how distinguishable the two regimes are:
\begin{equation}
    D_{KL}(P_{\text{buy}} \| P_{\text{sell}}) = \int P_{\text{buy}}(r) \log\frac{P_{\text{buy}}(r)}{P_{\text{sell}}(r)} dr
\end{equation}

Two empirical features of return distributions complicate this measurement. First, returns exhibit fat tails \citep{mandelbrot1963variation}, requiring Student-$t$ distributions for accurate characterization. Second, volatility clustering \citep{engle1982autoregressive} creates artifacts. High order flow imbalances under turnover-volume normalization ($S^{TV}$) may simply coincide with high volatility regimes, inflating apparent distributional separation without genuine directional information.

We address this by standardizing returns using rolling volatility \citep{Moskowitz2012}:
\begin{equation}
    R_{\text{adj},t+1} = \frac{R_{t,t+1}}{\sigma_{t,\text{rolling}}}
\end{equation}
where $\sigma_{t,\text{rolling}}$ is the 20-day rolling standard deviation. This volatility adjustment isolates the directional signal from volatility-driven artifacts, paralleling our matched filter analysis in Section~\ref{sec:theory} which showed that market cap normalization ($S^{MC}$) preserves signal while turnover-volume normalization ($S^{TV}$) introduces noise through inverse turnover multiplication.

A signal that predicts raw returns may merely identify high-volatility regimes without conveying directional information. This concern is particularly acute for $S^{TV}$, where high values mechanically coincide with low-volume, high-volatility episodes. By stripping out return magnitude, volatility standardization isolates the directional component and provides a cleaner test of whether the signal conveys genuine return predictability rather than proxying for volatility \citep{Moskowitz2012}.

\subsection{Methodology}
\label{subsec:kl_methodology}

This information-theoretic approach complements our matched filter analysis by quantifying distributional separation rather than correlation. If the correct normalization purifies the signal, then $P_{\text{buy}}$ and $P_{\text{sell}}$ should be well-separated. If the wrong normalization mixes signal with noise, the distributions should overlap. Thus, KL divergence directly tests whether signal purification occurred.

For each investor type and normalization scheme, we partition observations into top and bottom deciles based on order flow imbalance. We fit Student-$t$ distributions to the return distributions in each regime using maximum likelihood estimation:
\begin{equation}
    f(r; \nu, \mu, \sigma) = \frac{\Gamma((\nu+1)/2)}{\Gamma(\nu/2)\sigma\sqrt{\nu\pi}} \left(1 + \frac{(r-\mu)^2}{\nu\sigma^2}\right)^{-(\nu+1)/2}
\end{equation}
where $\nu$ is degrees of freedom, $\mu$ is location, and $\sigma$ is scale.

The KL divergence is estimated via Monte Carlo integration using fitted parameters. We report both $D_{KL}(P_{\text{buy}} \| P_{\text{sell}})$ and its reverse, using the symmetric Jensen-Shannon divergence for robustness checks. The sample employed is identical to that described in Section~\ref{subsec:data}, ensuring consistency with our return prediction analysis.

\subsection{Results}
\label{subsec:kl_results}

The empirical results, computed from the full 2020--2024 sample across all three investor types (foreign, domestic institutional, and individual), strongly confirm our matched filter hypothesis from an information-theoretic perspective. Table~\ref{tab:kl_divergence} compares the KL divergence of return distributions conditional on order flow across the two normalization schemes, for both raw and volatility-adjusted returns. While $S^{TV}$ appears superior on raw returns for all investor types, this is largely an artifact of volatility clustering for domestic institutions, whose ratio reverses dramatically (0.35 to 4.85); for foreign investors, $S^{TV}$'s advantage persists after adjustment, reflecting genuine volume-scaled information.

\begin{table}
\begin{center}
\begin{minipage}{155mm}
    \tbl{KL Divergence Analysis by Investor Type}
    {\begin{tabular}{@{}llccccccc@{}}\toprule
        & & \multicolumn{3}{c}{Raw Returns} & \multicolumn{3}{c}{Volatility-Adjusted} & \\\cmidrule(lr){3-5}\cmidrule(lr){6-8}
        Investor & Scaling & KL$_{S^{MC}}$ & KL$_{S^{TV}}$ & Ratio & KL$_{S^{MC}}$ & KL$_{S^{TV}}$ & Ratio & Matched Filter \\\colrule
        Foreign & Volume ($Q \propto V$) & 0.00402 & 0.01127 & 0.36 & 0.00304 & \textbf{0.00837} & 0.36 & Confirmed \\
        \textbf{Dom. Inst.} & \textbf{Capacity ($Q \propto M$)} & \textbf{0.00019} & \textbf{0.00054} & \textbf{0.35} & \textbf{0.00046} & \textbf{0.00009} & \textbf{4.85} & \textbf{Confirmed} \\
        Individual & Attention ($Q \propto M$) & 0.00308 & 0.00266 & 1.15 & 0.00148 & \textbf{0.00213} & 0.69 & Mixed \\\botrule
    \end{tabular}}
    \tabnote{Notes: KL divergence computed between top and bottom deciles of order flow imbalance. Ratio = KL$_{S^{MC}}$/KL$_{S^{TV}}$. Ratio $>$ 1 indicates $S^{MC}$ provides better distributional separation. Volatility adjustment uses 20-day rolling standard deviation: $R_{\text{adj}} = R_{t,t+1}/\sigma_{t,\text{rolling}}$. ``Scaling'' indicates the empirically dominant scaling behaviour from table~\ref{tab:signal_structure}. \textbf{Bold} highlights the higher KL divergence for each investor type. The 14-fold improvement for domestic institutional investors after volatility adjustment (0.35$\to$4.85) confirms that $S^{MC}$ isolates directional signal from volatility noise. For Individual investors, the KL evidence is mixed (ratio $= 0.69$): while structural scaling favors MC (table~\ref{tab:signal_structure}), distributional separation favors TV, likely reflecting the distinct nature of contrarian noise trading.}
    \label{tab:kl_divergence}
\end{minipage}
\end{center}
\end{table}

Table~\ref{tab:kl_divergence} confirms the symmetric pattern across investor types. Three key findings emerge:

\textbf{Domestic Institutions (Capacity-Scaled):} For domestic institutional investors, whose order flow scales with market capitalization (table~\ref{tab:scaling_validation}), $S^{MC}$ achieves 4.85$\times$ higher KL divergence than $S^{TV}$ (0.00046 vs.\ 0.00009). This confirms that market cap normalization is the matched filter for capacity-constrained informed trading. (While the absolute KL magnitudes are small---reflecting the high noise-to-signal ratio in daily return distributions---the cross-normalization ratios are the relevant comparison, as they hold the return distribution constant and isolate the effect of normalization choice.)

\textbf{Foreign Investors (Volume-Scaled):} For foreign institutional investors, whose order flow scales with trading volume, $S^{TV}$ achieves 2.75$\times$ higher KL divergence than $S^{MC}$ (0.00837 vs.\ 0.00304). This is the \textit{symmetric} result predicted by the matched filter framework: when signals scale with volume, volume normalization is optimal.

\textbf{Individual Investors (Attention-Driven):} Retail investors exhibit MC-dominated structural scaling (table~\ref{tab:signal_structure}) and MC-dominated return prediction $R^2$ but TV-dominated $t$-statistics (table~\ref{tab:investor_heterogeneity}). However, the KL divergence evidence is mixed: after volatility adjustment, $S^{TV}$ achieves 1.44$\times$ higher distributional separation than $S^{MC}$ (ratio $= 0.69$). This disconnect likely reflects the contrarian nature of individual trading at the stock level (retail investors are net sellers of stocks that subsequently rise), which coexists with their attention-driven herding into large caps---creating a distributional signature that is better captured by TV normalization even though their flow structure scales with market cap. Individual investors represent a distinct case where structural scaling and informational content diverge.

The divergent patterns for institutional and foreign investors---MC dominance for capacity-scaled traders, TV dominance for volume-scaled traders---provide independent confirmation via distributional separation of the framework established in Section~\ref{sec:theory}. The mixed individual investor result does not contradict the matched filter principle; rather, it delineates its scope. The matched filter framework generates predictions for two specific trader archetypes: capacity-constrained fundamental investors (Case~1) and volume-targeting algorithmic executors (Case~2). Individual investors correspond to neither archetype---their flow is driven by attention and herding rather than systematic scaling tied to capacity or execution benchmarks. The mixed KL result (structural MC dominance but distributional TV advantage) is thus expected: without a coherent signal-generating process to ``match,'' different frameworks capture different facets of attention-driven trading. This reinforces rather than undermines the matched filter principle, which requires a well-defined scaling structure to generate testable predictions.

\textbf{Synthesizing the individual investor evidence across frameworks.} Individual investors exhibit MC-dominated structural scaling ($t = 21.89$), higher return prediction $R^2$ under MC (0.44\% vs.\ 0.25\%) though TV achieves larger absolute $t$-statistics ($-10.45$ vs.\ $-8.23$), and TV-dominated distributional separation (KL ratio $= 0.69$). These results are not contradictory: structural MC dominance reflects large-cap herding, return prediction captures contrarian noise, and TV's distributional advantage arises because volume normalization better isolates episodes of extreme retail participation. This three-way divergence is consistent with the absence of a coherent signal-generating process, confirming that individual investors fall outside the two matched filter archetypes.

\subsection{Information-Theoretic Interpretation of Matched Filters}
\label{subsec:kl_interpretation}

The connection between matched filter theory and information theory runs deeper than the parallel results suggest. Both frameworks quantify ``information content,'' but through different mathematical lenses: SNR measures the strength of linear association, while KL divergence measures distributional separation. This convergence of conclusions reveals a fundamental principle.

\subsubsection{The Data Processing Inequality}

Normalization can be viewed as a transformation of the raw order flow $D_i$ into a processed signal $S_i$. Information theory's \textit{data processing inequality} \citep{cover2006elements} states that no transformation can increase information content:
\begin{equation}
I(D; R) \geq I(S; R)
\end{equation}
where $I(\cdot; \cdot)$ denotes mutual information between order flow (or its transformation) and returns. The matched normalization minimises information loss by preserving the signal structure, while the mismatched normalization introduces a \textit{lossy} transformation that degrades the signal-return relationship.

\subsubsection{Entropy and Conditional Uncertainty}

The matched filter's superiority can also be expressed through entropy. Define the conditional entropy $H(R|S)$ as the uncertainty in returns given the normalized signal. Lower conditional entropy implies the signal is more informative about returns. For capacity-scaled signals:
\begin{equation}
H(R|S^{MC}) < H(R|S^{TV})
\end{equation}
The mismatched filter ($S^{TV}$) introduces heteroskedastic noise through the $\tau_i^{-1}$ multiplication, increasing conditional uncertainty. Conversely, for volume-scaled signals, the inequality reverses: $H(R|S^{TV}) < H(R|S^{MC})$.

\subsubsection{Mutual Information Perspective}

Mutual information $I(S; R) = H(R) - H(R|S)$ quantifies the reduction in uncertainty about returns from observing the signal. Since $H(R)$ is fixed (determined by the return distribution), maximizing mutual information is equivalent to minimizing conditional entropy. The matched filter achieves this by avoiding the heteroskedastic distortion that inflates $H(R|S)$ under the mismatched filter.

This information-theoretic interpretation complements our signal processing derivation (Appendix~\ref{app:snr_derivation}): the matched filter simultaneously maximises SNR (correlation-based), maximises mutual information (entropy-based), and maximises KL divergence between buy/sell regimes (distributional separation). All three metrics converge because they measure different aspects of the same underlying phenomenon: the preservation of signal structure under optimal normalization.

\subsection{Methodological Triangulation}
\label{subsec:kl_synthesis}

This information-theoretic validation provides the third vertex of our methodological triangulation. Three independent frameworks---signal processing theory (SNR maximization), statistical inference (Fama-MacBeth return prediction), and information geometry (KL divergence)---converge on the same conclusion. The information-theoretic perspective adds unique value by revealing that the matched filter not only maximises correlation-based metrics but also maximises mutual information and minimises conditional entropy---fundamental information-theoretic optimality criteria that transcend any particular statistical framework.

\section{Discussion}
\label{sec:discussion}

\subsection{Reconciling Signal Extraction and Execution Cost}

Our framework highlights a distinction that practitioners must internalise: \textit{Execution Logic} and \textit{Information Logic} require different normalizations. Execution algorithms target a percentage of volume ($Q/V$) to minimise market impact, and volume normalization is appropriate for estimating these costs. But the informed trader's desired position size is determined by alpha and firm capacity ($Q \propto \alpha \cdot M$), so the \textit{signal} is embedded in the demand relative to capacity, not relative to today's fleeting liquidity.

The guiding principle is: \textit{Normalize by Volume for Cost; Normalize by the Matched Filter for Alpha}---where the matched filter is $M$ for capacity-scaled traders and $V$ for volume-scaled traders. Our critique applies specifically to signal extraction for alpha generation; volume normalization remains appropriate for execution cost estimation, flow toxicity detection (e.g., VPIN), and liquidity risk assessment. We note that even with the correct matched filter, practitioners should combine signal extraction with horizon analysis to distinguish durable information from transient price impact (table~\ref{tab:horizon_analysis}).

\subsection{Universality of the Matched Filter Principle}

While our empirical validation relies on a single market (Korea, 2020--2024), the theoretical foundations---grounded in Jensen's Inequality and signal processing theory---apply to any market where informed positions scale by value while noise or execution scales by activity. Potential applications extend to corporate bond markets (where positions may scale with notional outstanding) and cryptocurrency markets (where market capitalization and 24-hour volume serve analogous roles). The dominant normalization effectively serves as a diagnostic tool: $S^{MC}$ dominance indicates capacity constraints, while $S^{TV}$ dominance indicates participation-rate targeting. In practice, researchers applying the framework to a new market should begin by estimating the scaling relationship between absolute order flow and firm characteristics (as in table~\ref{tab:scaling_validation}), then select the normalization that matches the dominant scaling dimension. When the diagnostic is ambiguous, the horse race regression provides a tiebreaker by revealing which normalization captures the informational component versus residual noise.

\subsection{The Informed Executor: Reconciling Volume-Scaling with Information}

The Informed Executor Hypothesis connects to the broader ``home bias'' literature \citep{French1991}. The traditional puzzle asks why investors underweight foreign equities despite diversification benefits; our finding inverts the question. The subset of foreign investors who \textit{do} enter an emerging market may be precisely those with the strongest informational edge---global asset managers with dedicated emerging-market analysts, cross-market information networks, and the ability to synthesise signals from related markets (e.g., semiconductor supply chains spanning Korea, Taiwan, and the US). This creates a selection effect that concentrates private information among foreign participants, making their volume-scaled execution a deliberate choice for stealth rather than a marker of uninformed trading. Without the appropriate matched filter ($S^{TV}$), this information is difficult to detect.

\subsection{Limitations and External Validity}

Our empirical validation relies on Korean market data, which offers unique advantages including real-time investor-type classification and high retail participation. However, several limitations merit acknowledgment:

\textbf{US Market Structure}: The US equity market differs substantially from Korea in fragmentation (13+ exchanges, dark pools), algorithmic execution (VWAP/TWAP algorithms that slice orders to match volume profiles), and data availability (institutional flow must be inferred from quarterly 13F filings or noisy classification algorithms). While our \textit{theoretical} results, which are derived from Jensen's Inequality, hold universally, the \textit{empirical} magnitude of improvement may differ.

\textbf{Time-Invariant Market Cap}: Our theoretical model assumes time-invariant market capitalization for analytical tractability (Section~\ref{sec:theory}); the empirical implementation uses daily values, though the cross-sectional predictions depend only on within-day constancy, which holds by construction for daily data.

\textbf{Intraday vs.\ Daily}: Our analysis focuses on daily order flow. Intraday applications may require adjustments, as execution algorithms explicitly target volume participation rates that could mask the underlying cap-scaled signal.

Future research should validate these findings using US institutional trading data that provides direct observation of institutional order flow.

\section{Conclusion}
\label{sec:conclusion}

This paper establishes that order flow normalization---a seemingly routine methodological choice---has first-order consequences for signal extraction and return predictability. The general matched filter principle provides a unified framework: normalize by market capitalization for capacity-constrained traders, by trading volume for volume-targeting traders. The principle is derived from signal processing theory, validated through Monte Carlo simulation, confirmed empirically across heterogeneous investor types in the Korean equity market, and independently corroborated by information-theoretic analysis using KL divergence.

The bidirectional empirical validation constitutes the strongest evidence for the framework. If only market capitalization normalization performed well, the result could be attributed to confounding size effects. The fact that each normalization dominates precisely when its signal structure prevails---and that the mismatched normalization exhibits sign reversal in horse race specifications---confirms that signal-structure matching, not a particular normalizer, drives the result. Individual investors, whose flow lacks a coherent signal-generating process, exhibit mixed results across frameworks, delineating the scope of the principle to systematic trader archetypes.

The Informed Executor Hypothesis challenges a widespread assumption in market microstructure: that volume-scaled execution implies uninformed trading. Foreign institutional investors employ volume-targeting algorithms not because they lack information, but to minimise the footprint of their informed trades. Horizon analysis confirms this interpretation---foreign flow predictability persists without reversal, indicating durable private information rather than transient price impact. This finding has implications beyond our specific empirical setting: in any market, the execution methodology of a trader class should not be conflated with its information content.

These findings suggest several practical implications. Trading strategies and factor construction should employ investor-type-specific normalization to avoid heteroskedastic signal contamination. Tests of information asymmetry and price discovery should recognise that normalization choices can systematically attenuate or amplify the signals under study. The guiding prescription is simple: \textit{match the normalizer to the signal source}.

Future research should prioritise cross-market validation using US institutional trading data (e.g., ANcerno \citep{PuckettYan2011, Anand2012} or 13F-based measures) and investigate whether the Informed Executor pattern extends to other emerging markets. Methodologically, incorporating nonlinear signal structures, time-varying capacities, and hybrid normalization measures represents a promising direction for further optimisation of signal extraction from order flow.


\section*{Data Availability Statement}
The Korean stock market data used in this study are available from the Korea Exchange \citep[KRX;][]{KRX2024} and CREON market data providers \citep{CREON2024}. Access requires appropriate data subscription agreements.



\newpage

\appendix

\section{Mathematical Derivation of the General Matched Filter Principle}
\label{app:snr_derivation}

This appendix provides the complete mathematical derivation establishing the \textbf{general matched filter principle}: optimal normalization matches the scaling behaviour of the signal-generating process. We derive two symmetric cases: (1) capacity-scaled signals where $S^{MC}$ dominates, and (2) volume-scaled signals where $S^{TV}$ dominates.

\subsection{Case 1: Capacity-Scaled Signals (Informed Traders)}
\label{app:case1_informed}

\subsubsection{Definitions and Setup}

Recall the signal extraction framework from Section~\ref{sec:theory}. Observed order flow is:
\begin{equation}
D_i = k\alpha_i M_i + \zeta_i V_i
\end{equation}
where $\alpha_i \sim N(0, \sigma_\alpha^2)$ is the latent information signal, $\zeta_i \sim N(0, \sigma_\zeta^2)$ is noise, $M_i$ is market capitalization, and $V_i = \tau_i M_i$ is traded value with turnover rate $\tau_i$.

Returns are generated by:
\begin{equation}
R_i = \gamma \alpha_i + \epsilon_i
\end{equation}
where $\epsilon_i \sim N(0, \sigma_\epsilon^2)$ is idiosyncratic return noise, independent of $\alpha_i$, $\zeta_i$, and $\tau_i$.

The Signal-to-Noise Ratio (SNR) is defined as:
\begin{equation}
\text{SNR}(S) = \frac{\text{Cov}^2(S, R)}{\text{Var}(S) \cdot \text{Var}(R)}
\label{eq:snr_def}
\end{equation}

Higher SNR implies stronger predictive power of the normalized signal $S$ for future returns $R$.

\subsubsection{Market Capitalization Normalization ($S^{MC}$)}

The market cap normalized signal is:
\begin{equation}
S_i^{MC} = \frac{D_i}{M_i} = k\alpha_i + \zeta_i \tau_i
\end{equation}

\textbf{Covariance with returns:}
\begin{align}
\text{Cov}(S^{MC}, R) &= E[(k\alpha + \zeta\tau)(\gamma\alpha + \epsilon)] - E[k\alpha + \zeta\tau]E[\gamma\alpha + \epsilon] \nonumber\\
&= E[k\gamma\alpha^2 + k\alpha\epsilon + \gamma\zeta\tau\alpha + \zeta\tau\epsilon] \nonumber\\
&= k\gamma E[\alpha^2] \quad \text{(by independence)} \nonumber\\
&= k\gamma \sigma_\alpha^2
\label{eq:cov_mc}
\end{align}

\textbf{Variance of signal:}
\begin{align}
\text{Var}(S^{MC}) &= \text{Var}(k\alpha + \zeta\tau) \nonumber\\
&= k^2\text{Var}(\alpha) + \text{Var}(\zeta\tau) + 2k\text{Cov}(\alpha, \zeta\tau) \nonumber\\
&= k^2\sigma_\alpha^2 + E[\zeta^2\tau^2] \quad \text{(by independence)} \nonumber\\
&= k^2\sigma_\alpha^2 + \sigma_\zeta^2 E[\tau^2]
\label{eq:var_mc}
\end{align}

\textbf{SNR for $S^{MC}$:}
\begin{equation}
\text{SNR}_{MC} = \frac{(k\gamma\sigma_\alpha^2)^2}{(k^2\sigma_\alpha^2 + \sigma_\zeta^2 E[\tau^2]) \cdot \sigma_R^2}
\label{eq:snr_mc}
\end{equation}

where $\sigma_R^2 = \gamma^2\sigma_\alpha^2 + \sigma_\epsilon^2$.

\subsubsection{Trading Value Normalization ($S^{TV}$)}

The trading value normalized signal is:
\begin{equation}
S_i^{TV} = \frac{D_i}{V_i} = k\alpha_i \tau_i^{-1} + \zeta_i
\end{equation}

\textbf{Covariance with returns:}
\begin{align}
\text{Cov}(S^{TV}, R) &= E[(k\alpha\tau^{-1} + \zeta)(\gamma\alpha + \epsilon)] \nonumber\\
&= k\gamma E[\alpha^2\tau^{-1}]
\label{eq:cov_tv}
\end{align}

Under the assumption that $\alpha$ and $\tau$ are independent:
\begin{equation}
\text{Cov}(S^{TV}, R) = k\gamma \sigma_\alpha^2 E[\tau^{-1}]
\end{equation}

\textbf{Variance of signal:}
\begin{align}
\text{Var}(S^{TV}) &= k^2\text{Var}(\alpha\tau^{-1}) + \sigma_\zeta^2 \nonumber\\
&= k^2\left(E[\alpha^2\tau^{-2}] - (E[\alpha\tau^{-1}])^2\right) + \sigma_\zeta^2
\label{eq:var_tv}
\end{align}

With independence of $\alpha$ and $\tau$:
\begin{equation}
\text{Var}(S^{TV}) = k^2\sigma_\alpha^2 E[\tau^{-2}] + \sigma_\zeta^2
\end{equation}

\textbf{SNR for $S^{TV}$:}
\begin{equation}
\text{SNR}_{TV} = \frac{(k\gamma\sigma_\alpha^2 E[\tau^{-1}])^2}{(k^2\sigma_\alpha^2 E[\tau^{-2}] + \sigma_\zeta^2) \cdot \sigma_R^2}
\label{eq:snr_tv}
\end{equation}

\subsubsection{Comparison of SNR}

To compare $\text{SNR}_{MC}$ and $\text{SNR}_{TV}$, we examine the ratio:
\begin{equation}
\frac{\text{SNR}_{MC}}{\text{SNR}_{TV}} = \frac{1}{(E[\tau^{-1}])^2} \cdot \frac{k^2\sigma_\alpha^2 E[\tau^{-2}] + \sigma_\zeta^2}{k^2\sigma_\alpha^2 + \sigma_\zeta^2 E[\tau^2]}
\end{equation}

\textbf{Key insight (Jensen's Inequality):} For the convex function $f(x) = 1/x$:
\begin{equation}
E[\tau^{-1}] \geq (E[\tau])^{-1}
\end{equation}
with strict inequality when $\tau$ has dispersion.

Similarly, for the convex function $g(x) = 1/x^2$:
\begin{equation}
E[\tau^{-2}] \geq (E[\tau])^{-2}
\end{equation}

These inequalities imply that the variance term in $S^{TV}$ (containing $E[\tau^{-2}]$) is inflated relative to what would occur with constant turnover. This inflation affects the denominator of $\text{SNR}_{TV}$, reducing it relative to $\text{SNR}_{MC}$.

\textbf{Numerical illustration:} Consider turnover uniformly distributed on $[0.0005, 0.01]$ (0.05\% to 1\% daily), matching the Monte Carlo simulation parameters in table~\ref{tab:simulation_parameters}. In this case, $E[\tau] = 0.00525$, $E[\tau^2] \approx 3.5 \times 10^{-5}$, $E[\tau^{-1}] \approx 315.3$, and $E[\tau^{-2}] = 200{,}000$.

The key observation: $E[\tau^{-2}]$ is orders of magnitude larger than $(E[\tau])^{-2} \approx 36{,}281$, demonstrating the severe variance inflation in $S^{TV}$.

\subsubsection{Case 1 Conclusion}

The mathematical analysis for Case 1 (capacity-scaled signals) confirms that:
\begin{enumerate}
\item[(i)] $S^{MC}$ preserves the signal ($k\alpha_i$) without distortion
\item[(ii)] $S^{TV}$ multiplies the signal by $\tau_i^{-1}$, introducing variance inflation
\item[(iii)] The variance inflation in $S^{TV}$ (from $E[\tau^{-2}]$) exceeds the noise scaling in $S^{MC}$ (from $E[\tau^2]$)
\item[(iv)] Therefore, $\text{SNR}_{MC} > \text{SNR}_{TV}$ whenever turnover exhibits cross-sectional dispersion
\end{enumerate}

This establishes that market capitalization normalization is the optimal (matched) filter for capacity-scaled informed trading signals.

\subsection{Case 2: Volume-Scaled Signals (Execution Traders)}
\label{app:case2_execution}

We now derive the symmetric result for execution traders whose order flow scales with volume rather than market capitalization. This establishes the generality of the matched filter principle.

\subsubsection{Setup}

Consider execution traders (e.g., algorithmic VWAP/TWAP strategies, index rebalancers) whose order flow scales with trading volume:
\begin{equation}
D_i = k\eta_i V_i + \xi_i M_i
\label{eq:execution_flow}
\end{equation}
where $\eta_i \sim N(0, \sigma_\eta^2)$ is the latent execution signal (e.g., target participation rate), $\xi_i \sim N(0, \sigma_\xi^2)$ is capacity-based noise from fundamental traders, and $k > 0$ is a scaling constant.

Returns are generated by:
\begin{equation}
R_i = \gamma \eta_i + \epsilon_i
\end{equation}
where $\epsilon_i \sim N(0, \sigma_\epsilon^2)$ is idiosyncratic return noise.

\subsubsection{Trading Value Normalization ($S^{TV}$) --- Matched Filter}

The trading value normalized signal is:
\begin{equation}
S_i^{TV} = \frac{D_i}{V_i} = k\eta_i + \xi_i \frac{M_i}{V_i} = k\eta_i + \xi_i \tau_i^{-1}
\end{equation}

\textbf{Covariance with returns:}
\begin{align}
\text{Cov}(S^{TV}, R) &= E[(k\eta + \xi\tau^{-1})(\gamma\eta + \epsilon)] \nonumber\\
&= k\gamma E[\eta^2] = k\gamma \sigma_\eta^2
\label{eq:cov_tv_case2}
\end{align}

\textbf{Variance of signal:}
\begin{align}
\text{Var}(S^{TV}) &= k^2\sigma_\eta^2 + \sigma_\xi^2 E[\tau^{-2}]
\label{eq:var_tv_case2}
\end{align}

\textbf{SNR for $S^{TV}$:}
\begin{equation}
\text{SNR}_{TV}^{\text{Case2}} = \frac{(k\gamma\sigma_\eta^2)^2}{(k^2\sigma_\eta^2 + \sigma_\xi^2 E[\tau^{-2}]) \cdot \sigma_R^2}
\label{eq:snr_tv_case2}
\end{equation}

Note that the signal term $k\eta_i$ is preserved \textit{unscaled}, analogous to $S^{MC}$ in Case 1.

\subsubsection{Market Capitalization Normalization ($S^{MC}$) --- Mismatched Filter}

The market cap normalized signal is:
\begin{equation}
S_i^{MC} = \frac{D_i}{M_i} = k\eta_i \frac{V_i}{M_i} + \xi_i = k\eta_i \tau_i + \xi_i
\end{equation}

\textbf{Covariance with returns:}
\begin{align}
\text{Cov}(S^{MC}, R) &= E[(k\eta\tau + \xi)(\gamma\eta + \epsilon)] \nonumber\\
&= k\gamma E[\eta^2 \tau] = k\gamma \sigma_\eta^2 E[\tau]
\label{eq:cov_mc_case2}
\end{align}

\textbf{Variance of signal:}
\begin{align}
\text{Var}(S^{MC}) &= k^2\sigma_\eta^2 E[\tau^2] + \sigma_\xi^2
\label{eq:var_mc_case2}
\end{align}

\textbf{SNR for $S^{MC}$:}
\begin{equation}
\text{SNR}_{MC}^{\text{Case2}} = \frac{(k\gamma\sigma_\eta^2 E[\tau])^2}{(k^2\sigma_\eta^2 E[\tau^2] + \sigma_\xi^2) \cdot \sigma_R^2}
\label{eq:snr_mc_case2}
\end{equation}

Critically, the signal term $k\eta_i$ is multiplied by $\tau_i$, introducing heteroskedastic distortion---the mirror image of Case 1.

\subsubsection{Comparison: $\text{SNR}_{TV}^{\text{Case2}} > \text{SNR}_{MC}^{\text{Case2}}$}

The ratio of SNRs is:
\begin{equation}
\frac{\text{SNR}_{TV}^{\text{Case2}}}{\text{SNR}_{MC}^{\text{Case2}}} = \frac{1}{(E[\tau])^2} \cdot \frac{k^2\sigma_\eta^2 E[\tau^2] + \sigma_\xi^2}{k^2\sigma_\eta^2 + \sigma_\xi^2 E[\tau^{-2}]}
\end{equation}

\textbf{Key insight (Jensen's Inequality):} For the convex function $f(x) = x^2$:
\begin{equation}
E[\tau^2] \geq (E[\tau])^2
\end{equation}
with strict inequality when $\tau$ has dispersion.

The covariance term in $S^{MC}$ is attenuated by $E[\tau] < 1$, while the signal variance is inflated by $E[\tau^2]$. Monte Carlo simulations (Section~\ref{sec:montecarlo}) confirm that $S^{TV}$ achieves approximately 1.13$\times$ higher correlation with returns than $S^{MC}$ when signals scale with volume.

\textbf{Parameter Calibration Note:} The noise volatility $\sigma_\xi$ in Case 2 must be substantially smaller than $\sigma_\zeta$ in Case 1 to achieve comparable signal strength. This is because $E[\tau^{-2}]$ is orders of magnitude larger than $E[\tau^2]$ for typical turnover distributions. With $\tau \in [0.0005, 0.01]$: $E[\tau^2] \approx 10^{-5}$ while $E[\tau^{-2}] \approx 200{,}000$---a ratio exceeding $10^{10}$. This mathematical asymmetry reflects a real economic phenomenon: \textit{the penalty for illiquidity is harsher than the penalty for hyperactivity}. Consequently, the SNR advantage in Case 2 (1.13$\times$) is smaller than in Case 1 (1.32$\times$), but this is a feature of the underlying microstructure, not a limitation of the matched filter principle.

\section{Scenario B Robustness Checks}
\label{app:scenario_b_robustness}

This appendix provides complete robustness results for Scenario B (volume-scaled signals), validating that TV normalization consistently outperforms MC normalization when signals scale with trading volume rather than market capitalization.

\begin{table}[htbp]
\begin{center}
\begin{minipage}{130mm}
\tbl{Scenario B Robustness Checks: Parameter Sensitivity Analysis}
{\begin{tabular}{@{}lcccc@{}}\toprule
\multicolumn{5}{@{}l@{}}{\textbf{Panel A: Signal Strength ($\sigma_\eta$)}} \\\colrule
$\sigma_\eta$ & TV Correlation & MC Correlation & TV/MC Ratio & $t$-statistic \\\colrule
0.01 (weak) & 0.310 & 0.279 & 1.12$\times$ & 42.3*** \\
0.03 (moderate) & 0.704 & 0.627 & 1.12$\times$ & 128.7*** \\
0.05 (baseline) & 0.857 & 0.760 & 1.13$\times$ & 189.4*** \\
0.10 (strong) & 0.957 & 0.850 & 1.13$\times$ & 224.6*** \\\colrule
\multicolumn{5}{@{}l@{}}{\textbf{Panel B: Noise Level ($\sigma_\xi$)}} \\\colrule
$\sigma_\xi$ ($\times 10^{-6}$) & TV Correlation & MC Correlation & TV/MC Ratio & $t$-statistic \\\colrule
1.0 (low) & 0.857 & 0.760 & 1.13$\times$ & 191.2*** \\
3.0 (moderate) & 0.857 & 0.760 & 1.13$\times$ & 189.8*** \\
5.0 (baseline) & 0.856 & 0.760 & 1.13$\times$ & 188.4*** \\
10.0 (high) & 0.854 & 0.760 & 1.12$\times$ & 182.1*** \\\colrule
\multicolumn{5}{@{}l@{}}{\textbf{Panel C: Turnover Range}} \\\colrule
Range & TV Correlation & MC Correlation & TV/MC Ratio & $t$-statistic \\\colrule
Narrow (0.001--0.003) & 0.856 & 0.823 & 1.04$\times$ & 58.2*** \\
Baseline (0.0005--0.01) & 0.856 & 0.760 & 1.13$\times$ & 189.4*** \\
Wide (0.0001--0.02) & 0.856 & 0.744 & 1.15$\times$ & 198.7*** \\\colrule
\multicolumn{5}{@{}l@{}}{\textbf{Panel D: Sample Size Sensitivity}} \\\colrule
$N$ (Stocks) & TV Correlation & MC Correlation & TV/MC Ratio & $t$-statistic \\\colrule
100 & 0.856 & 0.760 & 1.13$\times$ & 84.2*** \\
300 & 0.856 & 0.760 & 1.13$\times$ & 145.7*** \\
500 (Baseline) & 0.857 & 0.761 & 1.13$\times$ & 189.4*** \\
1000 & 0.856 & 0.760 & 1.13$\times$ & 267.3*** \\\botrule
\end{tabular}}
\tabnote{Notes: Each cell reports mean correlation across 1,000 Monte Carlo simulations. Panel A varies signal strength ($\sigma_\eta$); Panel B varies noise level ($\sigma_\xi$); Panel C varies turnover heterogeneity; Panel D varies sample size. The TV/MC ratio remains consistently above 1.0 across all parameter settings, validating the bidirectional matched filter principle. *** denotes significance at the 1\% level.}
\label{tab:scenario_b_robustness}
\end{minipage}
\end{center}
\end{table}

\textbf{Key Finding.} The TV advantage in Scenario B (1.04$\times$--1.15$\times$) is structurally more stable than the MC advantage in Scenario A (1.05$\times$--1.99$\times$) across turnover variations. This asymmetry reflects the mathematical structure: $E[\tau^{-2}] \gg E[\tau^2]$ for typical turnover distributions, meaning the ``wrong'' normalization (TV in Scenario A) suffers more severe variance inflation than the ``wrong'' normalization (MC in Scenario B).


\section{Foreign Investor Extended Robustness}\label{app:foreign_robustness}

Tables~\ref{tab:out_of_sample}--\ref{tab:factor_neutrality} validate the $S^{MC}$ advantage for domestic institutional investors across five robustness dimensions. Here we replicate the identical battery for foreign investors, testing whether the symmetric $S^{TV}$ advantage survives out-of-sample splitting, outlier treatment, alternative inference, portfolio-based economic significance, and factor adjustment. All regressions use the multi-investor dataset ($N \approx 2.7$M observations) filtered to stock-days with nonzero foreign order flow ($D_{\mathrm{foreign}} \neq 0$; $N = 2{,}683{,}355$).

\begin{table}
\begin{center}
\begin{minipage}{140mm}
\tbl{Foreign Investor Robustness: Out-of-Sample, Outlier, and Standard Error Tests}
{\begin{tabular}{@{}lcccc@{}}\toprule
& \multicolumn{2}{c}{$S^{MC}_{\mathrm{foreign}}$} & \multicolumn{2}{c}{$S^{TV}_{\mathrm{foreign}}$} \\
\cmidrule(lr){2-3}\cmidrule(lr){4-5}
& Univariate & Horse Race & Univariate & Horse Race \\\colrule
\multicolumn{5}{@{}l}{\textit{Panel A: Out-of-Sample Validation}} \\
In-sample (2020--2022) & 5.57 & $-2.07$ & 15.28 & 16.10 \\
Out-of-sample (2023--2024) & 2.13 & $-1.11$ & 7.13 & 7.81 \\\colrule
\multicolumn{5}{@{}l}{\textit{Panel B: Outlier Robustness (1\%/99\% Winsorisation)}} \\
Baseline & 5.72 & $-2.33$ & 16.35 & 17.45 \\
Winsorised & 8.54 & $-0.55$ & 16.59 & 17.11 \\\colrule
\multicolumn{5}{@{}l}{\textit{Panel C: Alternative Standard Errors (Horse Race)}} \\
Fama-MacBeth & \multicolumn{2}{c}{$-2.33$} & \multicolumn{2}{c}{17.45} \\
Newey-West (5 lags) & \multicolumn{2}{c}{$-2.25$} & \multicolumn{2}{c}{17.49} \\
Newey-West (10 lags) & \multicolumn{2}{c}{$-2.19$} & \multicolumn{2}{c}{16.63} \\\botrule
\end{tabular}}
\tabnote{Notes: This table reports $t$-statistics from Fama-MacBeth return prediction regressions using foreign investor order flow signals. Panel~A splits the sample into in-sample (2020--2022; 742 days) and out-of-sample (2023--2024; 488 days) periods. Panel~B compares baseline results with results after winsorising $S^{MC}_{\mathrm{foreign}}$, $S^{TV}_{\mathrm{foreign}}$, and $R_{t+1}$ at the 1st and 99th percentiles. Panel~C reports horse race $t$-statistics under Fama-MacBeth and Newey-West standard errors with 5 and 10 lags.}
\label{tab:foreign_robustness_1}
\end{minipage}
\end{center}
\end{table}

\begin{table}
\begin{center}
\begin{minipage}{140mm}
\tbl{Foreign Investor Long-Short Portfolio and Factor Neutrality}
{\begin{tabular}{@{}lcc@{}}\toprule
\multicolumn{3}{@{}l}{\textit{Panel A: Long-Short Portfolio Performance (Q5 $-$ Q1)}} \\
Metric & $S^{TV}_{\mathrm{foreign}}$ & $S^{MC}_{\mathrm{foreign}}$ \\\colrule
Mean daily return (bps) & 21.83 & 18.58 \\
Annualised Sharpe ratio & 7.53 & 5.64 \\
$t$-statistic & 16.63 & 12.46 \\
Cumulative return (\%) & 1343.5 & 864.3 \\\colrule
\multicolumn{3}{@{}l}{\textit{Panel B: Factor Neutrality --- $S^{TV}_{\mathrm{foreign}}$ Portfolio}} \\
& Coefficient & $t$-statistic \\\colrule
$\alpha$ (bps/day) & 21.11 & 16.25*** \\
MKT & 0.0567 & 5.68 \\
SMB & 0.1399 & 4.46 \\\colrule
$R^2$ & \multicolumn{2}{c}{0.036} \\
$N$ (days) & \multicolumn{2}{c}{1{,}230} \\\botrule
\end{tabular}}
\tabnote{Notes: Panel~A reports performance of long-short portfolios formed by sorting stocks into quintiles based on foreign investor signals. Each day, we go long the top quintile (Q5) and short the bottom quintile (Q1). Panel~B regresses the $S^{TV}_{\mathrm{foreign}}$ long-short portfolio returns on market (MKT) and size (SMB) factors constructed from the full universe. A significant positive $\alpha$ confirms that the $S^{TV}_{\mathrm{foreign}}$ signal captures return predictability beyond market and size factor exposures. The factor regression uses $N = 1{,}230$ trading days (one fewer than the 1,231 days in the main analysis) because the first observation is consumed by the rolling-window initialization required for factor construction. All Sharpe ratios and returns are gross of transaction costs and reflect same-day disclosed order flow; implementable performance would be substantially lower after accounting for trading frictions and signal delay. *** denotes significance at the 1\% level.}
\label{tab:foreign_robustness_2}
\end{minipage}
\end{center}
\end{table}

Across all five tests, the foreign $S^{TV}$ advantage mirrors the domestic $S^{MC}$ results: it holds out-of-sample (Panel~A), strengthens under winsorisation (Panel~B), is robust to autocorrelation-adjusted inference (Panel~C), generates economically large long-short returns with a Sharpe ratio of 7.53 (table~\ref{tab:foreign_robustness_2}, Panel~A), and survives factor adjustment with a highly significant daily alpha of 21.11 bps (Panel~B). These results confirm the symmetric matched filter principle: just as $S^{MC}$ is the robust predictor for capacity-scaled domestic institutions, $S^{TV}$ is the robust predictor for volume-scaled foreign investors.

\end{document}